# Statistical and Economic Evaluation of Time Series Models for Forecasting Arrivals at Call Centers

Andrea Bastianin
University of Milan, Italy

Marzio Galeotti
University of Milan
IEFE Bocconi, Italy

Matteo Manera
University of Milan-Bicocca, Italy
Fondazione Eni Enrico Mattei, Milan

April, 2018

**Abstract:** Call centers' managers are interested in obtaining accurate point and distributional forecasts of call arrivals in order to achieve an optimal balance between service quality and operating costs. We present a strategy for selecting forecast models of call arrivals which is based on three pillars: (*i*) flexibility of the loss function; (*ii*) statistical evaluation of forecast accuracy; (*iii*) economic evaluation of forecast performance using money metrics. We implement fourteen time series models and seven forecast combination schemes on three series of daily call arrivals. Although we focus mainly on point forecasts, we also analyze density forecast evaluation. We show that second moments modeling is important both for point and density forecasting and that the simple Seasonal Random Walk model is always outperformed by more general specifications. Our results suggest that call center managers should invest in the use of forecast models which describe both first and second moments of call arrivals.

**Key Words:** ARIMA; Call center arrivals; Loss function; Seasonality; Telecommunications forecasting.

**JEL Codes:** C22, C25, C53, D81, M15.

# 1 Introduction

Hiring, staffing and scheduling are strategic decisions for the management of call centers, which represent a highly labor-intensive and large services industry, where human resources costs account for 60-70% of the operating budget (Gans et al., 2003). Point and density forecasts of call arrivals are a key input for choices relating to the acquisition and deployment of human resources, therefore accurate forecasts ultimately determine the ability of managers to achieve an optimal balance between service quality and operating costs (Akşin et al., 2007).

We present a novel strategy to select time series models of daily call arrivals that is based on three pillars: (*i*) flexibility of the loss function; (*ii*) statistical evaluation of forecast accuracy; (*iii*) economic evaluation of forecast performance using money metrics. We estimate fourteen time series models - including the Seasonal Random Walk (SRW) as a benchmark - that capture different key features of daily call arrivals Data of this sort are typically characterized by the presence of intra-weekly and intra-yearly seasonality, inter-day dependency (i.e. non-zero auto-correlation), over-dispersion (i.e. the variance of the arrival count per time period is larger than its expected value) and conditional heteroskedasticity. These features are shared by the three series we analyze in this paper, namely daily arrivals at call centers operated by an Italian electric utility and by two retail banks, one in the U.S. and the other in Israel.

The use of a flexible loss function and the implementation of statistical tests to rank and select forecasting models represent the first novelty the paper. In fact, in this strand of the literature, most studies only provide model rankings based on symmetric loss functions or informal forecast comparisons (see Ibrahim et al., 2016 for a survey). Since over–forecasting leads to over–staffing, hence to unnecessarily high operating costs, while under–forecasting results in under–staffing and in low service quality, the choice of the metric used to evaluate competing time series models depends on the preferences of the call center management, that are not necessarily well captured by a symmetric loss function. We rely on the loss function put forth by Elliott et al. (2005), that nests both symmetric and asymmetric loss functions as special cases and describes a wide range of call center managers' preferences.

The second novelty of this study is the translation of statistical measures of forecasting performance into money metrics. Money measures of performance are intimately related to the profit maximizing behavior of economic agents and should be considered by call center managers as more intuitive evaluation instruments to complement the information provided by loss functions and statistical tests (Leitch and Tanner, 1991).

Moreover, since it is well documented that combined forecasts often outperform forecasts generated by individual models (Timmermann, 2006), the third novelty of the paper is to implement seven forecast combination methods applied to five sets of models. Overall, we produce a total of 47 alternative forecasts.

We show that second moment modeling using Generalized Autoregressive Conditional Heteroskedasticity (GARCH) models is the preferred approach when forecasting daily call arrivals, which suggests that volatility modeling is useful. This result is robust, since it holds true not only for the call center managed by the Italian utility, but also for the two retail banks' arrival series. The simple SRW model is always outperformed by other, more general, easily implementable specifications, both for point and density forecasts. From the point of view of a call center manager, our findings imply that it is worth investing in the use of forecasting models which describe both the first and second moments of call arrivals.

Two papers are closely related to our work. The economic evaluation of the forecast accuracy we propose is similar in spirit to Shen and Huang (2008), although with some important distinctions. Our accuracy metric is the money a manager can earn, while their is the staffing level; moreover, we design a compensation scheme that incorporates asymmetric preferences, whereas they rely on a symmetric loss function; lastly, the forecast horizon is different. The second paper which is related to our work is Taylor (2008), since, to the best of our knowledge, this is the only contribution that evaluates a large number of time series models and forecast combination schemes. Compared with Taylor (2008), we do not deal with intra–day forecasts, but we evaluate the performance of models more thoroughly.

More generally, our paper belongs to the literature dealing with forecasts of daily call volumes (Andrews and Cunningham, 1995; Antipov and Meade, 2002; Bianchi et al., 1998; Mabert, 1985), but is also related with more recent studies focusing on density and intra–day predictions (Kim et al., 2012; Taylor, 2008, 2012; Tych et al., 2002; Weinberg et al., 2007).[1]

[1]There are at least two ways in which our results can be used to derive intra–daily forecasts. First, daily

The plan of the paper is as follows. Section 2 describes data, empirical methods and our approach to the economic evaluation of forecasts. Empirical results are discussed in Section 3. Extensions and robustness checks are presented in Section 4, while Section 5 concludes.

# 2 Data and methods

## 2.1 Data

We analyze three series of daily call arrivals received by call centers in different industries and countries. The main results of the paper are obtained from the arrival series of a call center managed by an anonymous Italian energy utility. Robustness checks in Section 4 extend the analysis using call arrival series at call centers operated by two anonymous retail banks, one located in Israel, the other in the U.S.. The number of observations is 749 for Italy, 361 for Israel and 893 for the U.S..[2] To save space, descriptive statistics, plots and statistical tests for the three series appear in an Appendix that is available from the authors upon request.

Being daily totals, the three series exhibit large count values, but with very different sample averages: the average daily call volume is approximately 1000, 31000, 45000 for the Israeli, the Italian and the U.S. call centers. Call arrivals for the Italian utility are equal to zero in correspondence of closing days associated with public holidays. These days are known in advance and are kept in the estimation sample by substituting the zeros with the number of calls recorded during the previous week. Forecasts for these days are subsequently set to zero. The rest of the series values are always strictly positive. The analysis of the coefficient of variation of the three series, calculated on their levels and on their log–transform, shows arrivals can be thought as part of a top–down approach (Gans et al., 2003), where forecasts of daily call arrivals are translated into hourly density forecasts by some simple methods, as illustrated in Channouf et al. (2007). Second, we can assume that the call center manager designs intra–day staffing schedules using a judgmental process based on his/her forecasts of daily totals.

[2]We thank Avi Mandelbaum for providing access to the retail banks data through the Technion Service Enterprise Engineering (SSE) Laboratory. See `http://ie.technion.ac.il/Labs/Serveng` for a detailed description of these series. The call arrivals for the Italian utility have been collected by the authors, and neither the name of company, nor the data can be disclosed.

that in all cases the logarithm seems to stabilize the variability of each series. Therefore all models, except count data specifications, will rely on log–transformed data.

Count data specifications include the Poisson and the Negative Binomial models. The Poisson model implies that the variance of call arrival volumes in each period is equal to its expected value during the same time frame. This property — known as equi–dispersion — is often not consistent with the features of call arrival data, which are likely to be over–dispersed, with variance larger than the mean. The Negative Binomial distribution includes the Poisson as a special case and allows for both under– and over–dispersion. A test of the null hypothesis of equi–dispersion, implemented along the lines put forth by Cameron and Trivedi (1990), provides evidence against the mean–variance equality. Moreover, estimates of the Negative Binomial model suggest that the three series are over-dispersed.[3]

The call center operated by the Italian energy utility, whose employees mainly help customers with invoicing problems, operates fourteen hours per day. A plot of this series appears in Figure 1 and suggests the presence of both daily and monthly seasonality. Specifically, the number of incoming calls decreases steadily from Monday to Sunday. Moreover, given the nature of the service provided by the Italian company, the intensity of calls varies with the season of the year. Lastly, even controlling for seasonality and autoregressive dynamics, the LM test for ARCH effects rejects the null hypothesis of homoskedasticity.

The call arrival series for the two banks also display day-of-the-week seasonality, but month–of–the–year seasonal factors seem less evident. Moreover, only the U.S. series exhibits conditional heteroskedasticity. On the contrary, for Israely call arrivals we cannot reject the null of homoskedasticity.[4]

[3] Details are reported in the Appendix

[4] We have also implemented a test of seasonality in the variance. Since the null hypothesis of no seasonality in the variance can be rejected for the U.S. series only, we have decided not to incorporate this type of seasonality in the forecasting models presented in our paper.

Figure 1: Daily call arrivals for the Italian call center

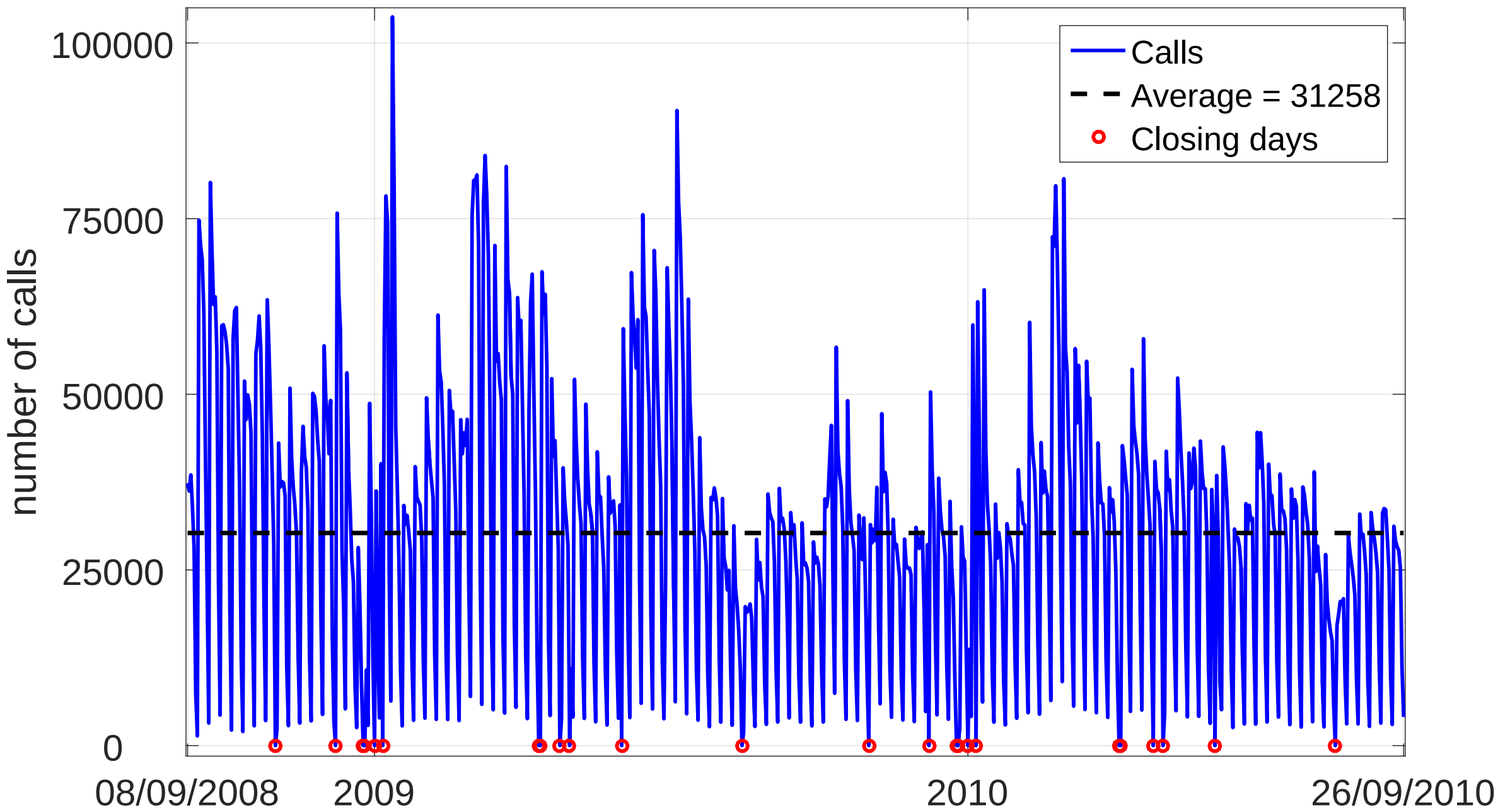

*Notes:* the figure shows the daily call arrival series (continuous line), its sample average (horizontal dashed line) and closing days (circles).

## 2.2 Forecasting models and methods

Given the nature of our series, all models have been chosen to capture different key features of the data, such as autocorrelation, seasonality, over–dispersion and conditional heteroskedasticity. The empirical specifications implemented in our analysis are shown in Table 1 and can be divided into three families.

The first family includes the SRW model, which is used as a benchmark, and a variety of time series models incorporating day–of–the–week dummies as exogenous variables. We focus on well established specifications, such as the Box–Jenkins Airline model, ARMAX and SARMAX models with and without GARCH effects, the Periodic Autoregressive (PAR) model and multiplicative Holt–Winters exponential smoothing. Although we are aware that the SRW model is simplistic, nonetheless we refer to this model as a benchmark, since it is widely used not only in this strand of the literature (see e.g. Taylor, 2008, 2012), but also in practical situations (see Mehrotra and Fama, 2003, p. 135).

The second family is formed by simple dynamic models for count data, which we think are reasonable specifications in the present context. In fact, the total number of calls arriving at call centers in a given time period is a count, which is usually modeled as a Poisson arrival process (Gans et al., 2003). The underlying assumption is that there is large population of potential customers, each of whom makes calls independently with a very low probability; that is, the total number of incoming calls in a given time period is approximately Poisson (Ibrahim et al., 2016). The Poisson modeling approach posits that the arrival rate is a deterministic function of the regressors. This implies that the variance of the arrival count per time period is equal to its expected value. Since in our case there is evidence of over–dispersion, we can deal with this feature of the data by assuming the arrival rate to be stochastic. We consider a Poisson distribution with Gamma–distributed arrival rate, which implies that the call arrival series follows a Negative Binomial distribution. Another advantage of this model is that it can also be used to produce density forecasts of the arrival rate. Following Jung and Tremayne (2011), who have shown that there is not a dominating modeling approach when forecasting with count data, we consider three specifications based on the Exponential, Poisson and Negative Binomial distributions.

We add to the first two families of standard models a third family of seasonal autoregressive specifications that have not been previously used to predict incoming calls (see Taylor, 2008, and references therein). In particular, a linear model with smoothly changing deterministic seasonality (i.e. the Time Varying Dummy AR, TVD–AR) is implemented to evaluate whether taking into account changes in the deterministic seasonal pattern of the series can improve the forecasting performance of daily call arrivals. As pointed out by Franses and van Dijk (2005), given that changes in technology, institutions, habits and tastes usually occur gradually, shifts of the seasonal pattern can be described with a smooth function (i.e. a logistic function) of the day–of–the–week dummy variables. The Multiplicative Error Model (MEM) — designed for dealing with non-negative variables such as volatility, volume and duration — has been introduced by Engle (2002) and shares many features with GARCH models, as well as with count data models and SARMAX specifications. Lastly, we forecast the series of incoming calls using a two–step modeling strategy. First, we remove from the series the changes in its sample average with a natural cubic spline function. Second, we apply a SARMAX model on the adjusted series, in order to capture the slowly–moving changes in the sample average of the series due to low–frequency seasonal factors.

Table 1: Summary of models

| id | Name | Dependent Variable | Explanatory Variables |
|---|---|---|---|
| $\mathcal{M}_0$ | Seasonal Random Walk | $Y_t$ | $Y_{t-7}$ |
| $\mathcal{M}_1$ | ARMAX | $y_t$ | AR(1), AR(7), AR(8), MA(1), $\mathbf{D}_t$ |
| $\mathcal{M}_2$ | ARMAX-GARCH(1,1) | $y_t$ | AR(1), AR(7), AR(8), MA(1), $\mathbf{D}_t$ |
| $\mathcal{M}_3$ | TVD-AR | $y_t$ | $\mathbf{D}_t$ |
| $\mathcal{M}_4$ | SARMAX | $y_t$ | AR(1), SAR(7), MA(1), SMA(28), $\mathbf{D}_t$ |
| $\mathcal{M}_5$ | SARMAX-GARCH(1,1) | $y_t$ | AR(1), SAR(7), MA(1), SMA(28), $\mathbf{D}_t$ |
| $\mathcal{M}_6$ | PAR(2) | $y_t$ | $y_{t-1}, y_{t-2}, \mathbf{D}_t$ |
| $\mathcal{M}_7$ | Airline | $\Delta \times \Delta_7 y_t$ | MA(1), SMA(8) |
| $\mathcal{M}_8$ | Poisson | $Y_t$ | $Y_{t-1}, \mathbf{D}_t$ |
| $\mathcal{M}_9$ | NegBin | $Y_t$ | $Y_{t-1}, \mathbf{D}_t$ |
| $\mathcal{M}_{10}$ | Exponential | $Y_t$ | $Y_{t-1}, \mathbf{D}_t$ |
| $\mathcal{M}_{11}$ | MEM | $y_t/\hat{y}_{SR,t}$ | $y_{t-1}/\hat{y}_{SR,t-1}, \mathbf{D}_t$ |
| $\mathcal{M}_{12}$ | Spline-SARX | $y_t/\hat{y}_{LR,t}$ | AR(1), SAR(7), $\mathbf{D}_t$ |
| $\mathcal{M}_{13}$ | Holt-Winters | $y_t$ | Multiplicative |

*Notes*: $Y_t$ is the number of incoming calls; $y_t \equiv \log Y_t$; $\mathbf{D}_t$ is a vector of dummies, one for each day of the week; $\Delta_k = (1 - L^k)$ where $L$ is the lag operator; $\hat{y}_{SR,t}$ denotes fitted values from the regression of $y_t$ on the vector of dummies; $\hat{y}_{LR,t}$ denotes fitted values from the interpolation of $y_t$ with a natural cubic spline, with the number of knots that equals the number of months in the sample; ARMA and seasonal ARMA terms are denoted as AR(.), MA(.), SAR(.) and SMA(.), where the number in brackets represents their order; "multiplicative" indicates that forecasts from $\mathcal{M}_{13}$ are obtained with the multiplicative Holt-Winters exponential smoothing, see Gardner (2006) for details.

Model estimation and forecasting is carried out recursively, that is the estimation sample expands by including a new observation at each iteration. The first iteration relies on an estimation sample of $R = 371$ days. The forecast horizon, $h$, ranges from one day to one month (28 days). The recursive scheme implies that the number of predictions, $P_h$, $h = 1, \ldots, 28$, varies from $P_1 = 378$ to $P_{28} = 351$.

For each model, the selection of the optimal specification is performed only once, using the sample of data pertaining to the first iteration of the estimation–forecasting scheme. We have selected most of the specifications in Table 1 using both the Schwarz Information Criterion (SIC) to choose the optimal number of lags (or, for some models, SARMA terms) and a stepwise regression approach.[5] Models are also subject to non-rejecting at 5% critical level the null hypothesis of absence of error autocorrelation tested with a Lagrange Multiplier test against the alternative hypotheses of first-to-eighth order error autocorrelation. For ARMA($p$,$q$) models we set $p^{\max}, q^{\max} = 28$; for seasonal AR($k$) and MA($l$) terms, we tried

[5]The stepwise approach starts with a model including all the explanatory variables selected with the SIC: Then it is repeatedly applied until all variables included in the final specification are significant at the 5% critical level.

$k^{\max}, l^{\max} = 7, 14, 21, 28$.

Some models include a GARCH component because squared residuals from ARMAX and SARMAX specifications display some neglected dynamics. Moreover, the inclusion of a GARCH equation can help us to shed light on the usefulness of modeling second moments for forecasting call arrivals. As for count data models, one lag of the dependent variable is proved to be enough to remove most of the autocorrelation in the residuals. A parsimonious specification of the dynamics involved in our count data models allows us to focus the forecast comparison on the role played by the distributional assumptions.

## 2.3 Forecast combination methods

Since combined forecasts are often found to outperform individual models (see Timmermann, 2006), we implement several combination schemes to predict call arrivals.

As shown in the upper panel of Table 2, our fourteen models are collected into five groups, which do not include the SRW benchmark model. The first group, $\mathcal{G}_1$, is the largest, for it excludes the Holt-Winters exponential smoothing only. The second group of models, $\mathcal{G}_2$, includes two ARMAX specifications and the TVD-AR model. Group $\mathcal{G}_3$ differs from $\mathcal{G}_4$ in that the latter excludes the PAR model from the set containing ARMAX, TVD-AR and SARMAX models. The last group, $\mathcal{G}_5$, is composed by time series models for count data.

We combine forecasts from these five groups of models with average, trimmed average, median, minimum, maximum and Approximate Bayesian Model Averaging (ABMA) combining schemes (see the lower panel of Table 2). All these methods have a feature in common: they do not require a set of out–of–sample observations, hence they can be used in real–time by the forecast user.

Stock and Watson (2004) have shown that simple combining methods such as the average, the trimmed average and the median work well in macroeconomic forecasting. If compared to the simple average method, both the median forecast approach and the trimmed average combination method (that excludes the highest and the lowest forecasts) reduce the impact of individual outlying forecasts. The maximum and minimum combination methods are used to represent two opposite situations, that is a manager who is either adverse to

under–staffing, or a manager trying to minimize labor costs and who is not subject to any kind of service level agreement.

Table 2: Groups of models and combining methods

| id | Models |
|---|---|
| $\mathcal{G}_1$ | $\mathcal{M}_i$ for $i = 1, ..., 12$ |
| $\mathcal{G}_2$ | $\mathcal{M}_1, \mathcal{M}_2, \mathcal{M}_3$ |
| $\mathcal{G}_3$ | $\mathcal{M}_1, \mathcal{M}_2, \mathcal{M}_3, \mathcal{M}_4, \mathcal{M}_5, \mathcal{M}_6$ |
| $\mathcal{G}_4$ | $\mathcal{M}_1, \mathcal{M}_2, \mathcal{M}_3, \mathcal{M}_4, \mathcal{M}_5$ |
| $\mathcal{G}_5$ | $\mathcal{M}_8, \mathcal{M}_9, \mathcal{M}_{10}$ |
| **Method** | **Description** |
| Average ($c_1$) | $f^{c1}_{\mathcal{G}_i,t} = \frac{1}{M_{\mathcal{G}_i}} \sum_{m=1}^{M_{\mathcal{G}_i}} f_{m,t}$ |
| Trimmed Average ($c_2$) | $f^{c2}_{\mathcal{G}_i,t} = \frac{1}{M_{\mathcal{G}_i}-2} \sum_{m=1}^{\left(M_{\mathcal{G}_i}-2\right)} f_{m,t}$ |
| Median ($c_3$) | $f^{c3}_{\mathcal{G}_i,t} = \text{median}\left(\mathbf{f}_{\mathcal{G}_i,t}\right)$ |
| Min ($c_4$) | $f^{c4}_{\mathcal{G}_i,t} = \max\left(\mathbf{f}_{\mathcal{G}_i,t}\right)$ |
| Max ($c_5$) | $f^{c5}_{\mathcal{G}_i,t} = \min\left(\mathbf{f}_{\mathcal{G}_i,t}\right)$ |
| ABMA-SIC ($c_6$) | $f^{c6}_{\mathcal{G}_i,t} = \sum_{m=1}^{M_{\mathcal{G}_i}} w^{c_j}_{m,t} f_{m,t}$ |
| ABMA-AIC ($c_7$) | $f^{c7}_{\mathcal{G}_i,t} = \sum_{m=1}^{M_{\mathcal{G}_i}} w^{c_j}_{m,t} f_{m,t}$ |

*Notes*: models are described in Table 1. $f^{c_j}_{\mathcal{G}_i,t}$ denotes the forecast at time $t$ obtained with combining method $j$ on $\mathcal{G}_i$, for $j = 1, \dots, 7$ and $i = 1, \dots, 5$; $f_{m,t}$ is the forecast at time $t$ from model $m$, for $m = 1, \dots, M_{\mathcal{G}_i}$, where $M_{\mathcal{G}_i}$ is the number of models in the $i$-th group; $\mathbf{f}_{\mathcal{G}_i,t}$ is a $\left(M_{\mathcal{G}_i} \times 1\right)$ vector of forecasts from models in $\mathcal{G}_i$; Approximate Bayesian Model Averaging (ABMA) uses weights, $w^{c_j}_{m,t} = \frac{\exp\{\zeta_{m,t}\}}{\sum_{m=1}^{M_{\mathcal{G}_i}} \exp\{\zeta_{m,t}\}}$, where $\zeta_{m,t} = \text{IC}_{m,t} - \max\left(\mathbf{IC}_{\mathcal{G}_i,t}\right)$, for $m = 1, ..., M_{\mathcal{G}_i}$, $j = 6, 7$, $i = 2, ..., 5$ and IC = SIC, AIC; the $(\mathcal{M}_{\mathcal{G}_i} \times 1)$ vector, $\mathbf{IC}_{\mathcal{G}_i,t}$, contains the IC of models in the $i$-th group; combining method $c_6$ is based on the SIC, while $c_7$ uses the AIC; both exclude $\mathcal{G}_1$ from ABMA.

ABMA, successfully applied to macroeconomic forecasting by Garratt et al. (2003), uses the SIC and the Akaike Information Criterion (AIC) to approximate the posterior probability of individual models. ABMA is applied only to models whose dependent variable is expressed with a common unit of measure, i.e. to all groups of models, $\mathcal{G}_1$ excluded.

## 2.4 Statistical measures of forecast accuracy

The need for a loss function that can be either symmetric or asymmetric arises in many economic and management problems. A flexible loss functions serves two purposes. First, it allows to assign a different cost to positive and negative forecast errors. Second, it helps the call center manager and the professional forecaster to decide the shape of the loss and to use the same metric of predictive performance. The statistical evaluation of forecasts is based on the loss function put forth by Komunjer and Owyang (2012) that have proposed a multivariate generalization of the loss function due to Elliott et al. (2005, 2008). The strength of these metrics is to encompass a variety of symmetric and asymmetric loss functions that are often used in empirical applications.

Let $f_{i,t+h|t}$ be the $h$–step ahead forecast of $Y_{t+h}$ issued with model $i$ ($\mathcal{M}_i$), where $h$ is the forecast horizon. The corresponding forecast error is $u_{i,t+h|t} = Y_{t+h} - f_{i,t+h|t}$. From now on we drop the model subscript $i$ for ease of notation. Let us consider a set of $(P \times 1)$ vectors of forecasts errors, one for each forecast horizon: $\mathbf{u}_{(h)} = [u_{(h)1}, \ldots, u_{(h)P}]'$ for $h = 1, \ldots, H$, where $H$ is the maximum forecast horizon (i.e. 28 days in our case) and $P$ is the size of the evaluation sample. These vectors form the rows of a $(H \times P)$ matrix: $\mathbf{u} = [\mathbf{u}_{(1)} \ldots \mathbf{u}_{(H)}]'$, where $\mathbf{u}_p$ for $p = 1, \ldots, P$ indicates its $p$-th column.

With this notation, we can write the multivariate loss function of Komunjer and Owyang (2012) as:

$$\mathcal{L}_p\left(\mathbf{u}_p; \rho, \boldsymbol{\tau}\right) = \left(\left\|\mathbf{u}_p\right\|_\rho + \boldsymbol{\tau}'\mathbf{u}\right)\left\|\mathbf{u}_p\right\|_\rho^{\rho-1} \tag{1}$$

where $\left\|\mathbf{u}_p\right\|_\rho \equiv \left(\sum_{h=1}^{H}\left|u_{(h)p}\right|^\rho\right)^{\frac{1}{\rho}}$ and $-1 \leq \tau \leq 1$. The shape of the function is determined by $\rho > 0$ and $\boldsymbol{\tau}$; we assume that all the elements of $\boldsymbol{\tau}$ are equal to $\tau$, so that the degree of asymmetry is the same at all forecast horizons $h$. The loss is symmetric and additively separable for $\tau = 0$ and includes some special cases. When $\tau = 0$ and $\rho = 2$, we obtain the trace of the Mean Squared Error (MSE) loss function, while for $\tau = 0$ and $\rho = 1$ equation (1) reduces to the trace of the Mean Absolute Error (MAE) loss function. In both cases, symmetry ensures the multivariate loss to be additively separable in univariate losses. On the contrary, when $\tau \neq 0$ the multivariate loss function does not correspond to the sum of individual losses.

We also evaluate the forecasts separately at different horizons, so that $H = 1$ and $\mathbf{u} = \mathbf{u}'_{(1)} = [u_{(1)1}, \ldots, u_{(1)P}]$. In this case we can simplify the notation further and write $\mathbf{u} = [u_1, \ldots, u_P]$. Equation (1) reduces to the univariate flexible loss of Elliott et al. (2005):

$$\mathcal{L}_p\left(u_p; \rho, \tau\right) \;=\; 2\left[\phi + (1 - 2\phi) \times I(u_p < 0)\right] \times |u_p|^\rho \tag{2}$$

where $I(.)$ is the indicator function and $\tau = 2\phi - 1$. This loss function is asymmetric for $\phi \neq 0.5$ ($\tau \neq 0$). Over–forecasting (negative forecast errors) is costlier than under–forecasting for $\phi < 0.5$ ($\tau < 0$). On the contrary, when $\phi > 0.5$ ($\tau > 0$), positive forecast errors (under-prediction) are more heavily weighed than negative forecast errors (over-prediction). Equation (2) nests several special cases. For $\rho = 1$, the lin–lin loss is obtained. Moreover, when $\phi = 0.5$ ($\tau = 0$), the function is symmetric and boils down to the MAE. For $\rho = 2$ we get the quad-quad loss, that coincides with the MSE loss for $\phi = 0.5$ ($\tau = 0$).[6]

For each model and combination method, we produce a total of 28 forecast error series, one for each forecast horizon, $h$. Given that presenting detailed results for each $h$ is not a viable option, we rank models according their overall performance using the multivariate loss in equation (1), while we rely on equation (2) to evaluate forecast performance at horizon $h = 1, 7, 28$. Hereafter we fix $\rho = 2$. In this case, with a common $\tau$ for all $h = 1, \ldots, H$, non-negativeness of $\mathcal{L}_p$ requires $|\tau| < \frac{1}{\sqrt{H}}$. Thus, for $H = 28$, we need $|\tau| < 0.18$ (since $\phi = 0.5 \times (\tau + 1)$, which implies $\phi \in (0.41, 0.59)$). For this reason, we vary the asymmetry parameter across the following set of values: $\phi = \{0.42, 0.50, 0.58\}$ (i.e. $\tau = \{0.16, 0.00, -0.16\}$). When $\phi = 0.5$, the ranking is equivalent to the MSE ranking in the univariate case and coincides with the trace of MSE ranking in the multivariate case. When $\phi = 0.42$, over–forecasting is costlier than under–forecasting, and vice versa for $\phi = 0.58$.[7]

## 2.5 Monetary measures of forecast accuracy

For the economic evaluation of forecasts, the magnitude of the asymmetry is not relevant. In other words, it would not be realistic to state that a practitioner has a clear opinion on the parameters of the loss function. On the contrary, we believe that it is crucial to take a stance on the direction of the asymmetry. An asymmetric loss function might be required when firms sign a contract with an external organization that manages a call center. These contracts often involve some sort of Service Level Agreement (SLA), which defines quality standards that the outsourced call centers should meet. For instance, the widely applied

[6]Optimal forecasts from the lin–lin and quad-quad loss are conditional quantiles and expectiles, respectively (Newey and Powell, 1987; Koenker and Bassett, 1978; Gneiting, 2011). See Bastianin et al. (2014) for the use of expectiles in a forecast horse–race.

[7]Additional details about flexible loss functions and their graphical representations appear in the Appendix.

80/20 SLA implies that eighty percent of the incoming calls must be answered within twenty seconds (Stolletz, 2003). SLAs not only set quality standards, but also dictate penalties for failing to meet contractual terms (Milner and Olsen, 2008). The presence of penalty fees might render a reasonable degree of over–forecasting less costly than the same amount of under–prediction, possibly because workers becoming free at short notice might be assigned to meetings and training (Taylor, 2008). Anecdotal evidence suggests that managers facing a SLA do prefer over–forecasting in order to minimize their chances of incurring a penalty.[8]

*Assumptions.* We assume that the call center operates under a 80/20 SLA and is subject to a fee when the waiting time exceeds a given threshold. Therefore, a reasonable degree of over–staffing is less costly than the same amount of under–staffing. Moreover, we assume that at day $t$ the manager uses his/her forecast of inbound calls for day $t+1$ in an algorithm determining the number of agents ($n_t$) needed to comply with the company's SLA.

The algorithm used to staff the call center is the Erlang–C queuing model. We assume that the average call duration is three minutes and that the call center is open fourteen hours a day. Notwithstanding its well–known limitations, the Erlang–C model is widely used in practice, possibly because of its simplicity (Akşin et al., 2007; Gans et al., 2003). Moreover, as shown by Jongbloed and Koole (2001), even when the assumptions underlying the Erlang–C algorithm are not valid (e.g. in the presence of over-dispersion in the data), it can still be used to obtain reliable interval estimates of the optimal service level.

[8]Roubos et al. (2012) observe that *"A higher than necessary service level is generally not a problem, but managers might be penalized for failing to meet the target in too many periods. To this end some managers deliberately opt for a higher expected service level or a lower target time in order to meet the original target with higher likelihood"*. Similarly, Thomas (2005) reports that *"(...), some managers facing contractual service–level commitments (...) "overshoot" to improve their chances of success. That is, if they need to exceed a 95% fill rate to avoid penalty, they might set the stock level based on a 97% long-run fill rate."* However, as emphasized by Tran and Davis (2011): *"Some scholars conclude that moderate under–staffing has positive effects on outcomes, while others prove that slight over–staffing performs better. They all agree that, however, both great over–staffing and extreme under–staffing conditions have negative effects on organization performance."* These comments also support our choices of restricting the degree of asymmetry of the loss function within a relatively short interval not only to ensure that it is non-negative, but also to avoid that the loss function could represent too extreme preferences toward under– or over–staffing. In fact, empirical estimates indicate that the degree of asymmetry of economic agents is not very large (Elliott et al., 2005).

Table 3: Multinomial payoff scheme

| $k$ | lower bound ($LB_k$) | upper bound ($UB_k$) | bonus / penalty (Euro) |
|---|---|---|---|
| 1 | 0.00 | 0.80 | -10 |
| 2 | 0.80 | 0.90 | -5 |
| 3 | 0.90 | 0.95 | -2.5 |
| 4 | 0.95 | 1.05 | 10 |
| 5 | 1.05 | 1.10 | -1.25 |
| 6 | 1.10 | 1.20 | -2.5 |
| 7 | 1.20 | $\infty$ | -10 |

*Notes*: the multinomial compensation scheme implies that at time $t$ the manager gets a bonus $b_t = 10$ Euro if $n_t^* \times LB_4 \leq n_t < n_t^* \times UB_4$.

Let the manager's daily payoff, $W_t$, be the sum of a fixed $F$ and a variable part, $v_t$, that is $W_t = F + v_t$. For the fixed part of the payoff, we impose that the call center manager earns on average 1200 Euro for 28 working days, that is $F = \lceil 1200/28 \rceil = 43$ Euro per day. The variable part of the payoff, $v_t$, depends on the manager's ability to staff the call center, $d_t$, that is $v_t \equiv v_t(d_t)$. $d_t$ is evaluated ex-post and it is defined as a function of the distance between his decision, $n_t$, and the optimal number of agents $n_t^*$, where $n_t^*$ is calculated using the realized number of incoming calls as input to the Erlang–C model.

The company relies on the compensation scheme displayed in Table 3. We have designed compensations so as to penalize under–staffing more heavily than over-staffing, assuming that the company's objective is to maximize customer satisfaction. Moreover, we assume that the company's compensation policy implies symmetry of over- and under–staffing if forecast errors of both signs exceed a given threshold. At the end of the forecasting sample, whose length is $P$, the manager's payoff will be: $\pi = P \times F + \sum_{t=1}^{P} v_t$.

Following Dorfman and McIntosh (1997), we assume that the manager has a negative exponential utility function $U(\pi) = 1 - \exp(-\lambda\pi)$, where $\lambda$ represents the manager's absolute risk aversion coefficient. Notice that, for the negative exponential utility function, $\lambda^{-1}$ describes the willingness to lose. Depending on $d_t$, the manager can either get a bonus ($b_t$), or be subject to a maximum penalty ($p_t$) of 10 Euro. Given that each day he/she can lose 20 Euro at most, $\lambda$ is varied according to the formula $\lambda = [j \times P \times (b_t + p_t)]^{-1}$, where $j = 0.1, 0.5, 0.7$ denotes a percentage of the variable part of the payoff. This implies that the willingness to lose, $\lambda^{-1}$, can take on the following values, expressed in Euros:

$\lambda^{-1} = \{732, 3660, 5124\}$. If the manager could always get the bonus, the total payoff would be $P \times (F+10) = 18603$ Euro, where $F = 43$ Euro and $P = 351$ days. Therefore, the values of the willingness to lose are equivalent to 0.4%, 19.7% and 27.5% of the total payoff.

The end–of–period expected utility is $EU(\pi) = 1 - M_{\pi}(-\lambda)$, where $M_{\pi}(-\lambda)$ is the Moment Generating Function (MGF) (see Collender and Chalfant, 1986; Elbasha, 2005; Gbur and Collins, 1989). This result and our compensation scheme allow to calculate the expected utility using Maximum Likelihood estimates of the multinomial MGF.[9]

*The economic value of information.* The economic value of information of a set of forecasts can only be determined with reference to an alternative set of forecasts. Following Dorfman and McIntosh (1997), we define the value of perfect information as the value of a model that generates perfect forecasts, that is $n_t = n_t^*$, $\forall t$.

If the manager could purchase this model, he/she would face no risk and the payoff distribution would be a single point at $\pi^* = \max(\pi)$. The lack of risk (i.e. $var(\pi^*) = 0$) implies that the value of perfect information, $V^*$, is simply the payoff obtainable from the perfect forecast, in other words $V^* = \pi^*$. Given that a forecast can be "consumed" only in discrete quantities, the expected marginal utility of the forecast, $MU$, equals its expected utility. Since in equilibrium the price ratio of two goods is equal to their marginal rate of substitution, the value of the $i-th$ forecasting model, $\mathcal{M}_i$, is the solution of: $V_i/V^* = MU_i/MU^*$. Solving for $V_i$, using $MU_i = EU_i$ and $V^* = \pi^*$, yields (see Dorfman and McIntosh, 1997):

$$V_i = \frac{\pi^* EU_i}{EU(\pi^*)} \tag{3}$$

Equation (3) can be used to define the incremental value of information of the forecast model $\mathcal{M}_i$ with respect to model $\mathcal{M}_0$ as:

$$\Delta V_i \equiv V_i - V_0 \tag{4}$$

Therefore, $\mathcal{M}_i \succ \mathcal{M}_0$, if $\Delta V_i > 0$ or, equivalently, if $V_i > V_0$.

[9] The MGF of a multinomially distributed random variable is equal to $\left(\sum_{k=1}^{r} p_k e^{t_k}\right)^P$. An estimate of the probability $p_k$ can be calculated as: $\hat{p}_k = \sum_{t=1}^{H} I(n_t \in CI_k)/H$, where $CI_k$ for $k = 1, ..., 7$, denotes the naive confidence interval in Table 3.

*The certainty equivalent.* An alternative money metric of forecast accuracy of model $\mathcal{M}_i$ is the certainty equivalent (CE), which is defined as the value $\widetilde{\pi} \equiv CE_i$ that solves $U(\widetilde{\pi}) = EU(\pi_i)$:

$$CE_i = -\frac{1}{\lambda} \log \left[1 - EU\left(\pi_i\right)\right] \tag{5}$$

We can state that $\mathcal{M}_i \succ \mathcal{M}_j$ if $CE_i > CE_j$. The CE can also be used to determine the maximum amount of money the manager is willing to pay in order to switch from model $\mathcal{M}_i$ to model $\mathcal{M}_j$.

We assume that the manager can choose between using the naive SRW forecast ($\mathcal{M}_0$) for free, or buying model $\mathcal{M}_i$ from an expert. Moreover, let us assume that buying the forecast model $\mathcal{M}_i$ costs $\delta_i$, where $\delta_i$ represents a fraction of the payoff the manager would get from the naive forecast, that is $\delta_i \equiv \theta\pi_0$ with $0 < \theta < 1$. The fraction of the payoff deriving from the naive model that the manager is willing to pay to use forecast model $\mathcal{M}_i$ can be written as:

$$\delta_i = CE_i - CE_0 \tag{6}$$

or, equivalently, as: $\theta = (CE_i - CE_0)/\pi_0$.

*Linking monetary and statistical measures of forecast accuracy.* Given the assumptions illustrated in the previous sections, when the statistical loss function is parametrized so as to penalize under–prediction more heavily than over–prediction (i.e. $\rho = 2$, $\phi = 0.58$), the economic evaluation of forecasts in economic terms becomes a useful alternative to the evaluation based on standard statistical metrics. On this respect, Dorfman and McIntosh (1997) and Leitch and Tanner (1991) have shown that money metrics of forecasting performance, such as the value of information and the CE, are more closely related to the profitability of forecasts than the traditional summary statistics based on loss functions.

# 3 Empirical results

Our main results can be summarized as follows:

- The benchmark SRW model is outperformed by relatively more general, easily imple-

mentable, methods.

- Modeling second moments is useful, as the addition of a GARCH component to ARMAX and SARMAX models improves their performance. Moreover, the SARMAX–GARCH model is among the best performing specifications.

- Combining forecasts from different models often outperforms the forecasts obtained from single models.

- ABMA is the preferred forecast combination method.

- The economic and statistical evaluation of models and combining methods deliver consistent results. The best individual forecasts are those involving second–moment modeling, while the best combining method is ABMA, based on a group of models that includes models with a GARCH component.

- The degree of complexity of the ABMA forecasting method is higher than producing forecasts with a single model. Then, when over-forecasting is less heavily penalized than under–forecasting, the economic evaluation approach identifies the SARMAX–GARCH model as the best option.

## 3.1 Statistical measures

Model rankings using univariate and multivariate flexible losses are presented in Table 4. The univariate loss rankings for forecast horizons at one day ($h = 1$), one week ($h = 7$) and one month ($h = 28$) are shown in columns 2-9, while rankings based on the multivariate loss for $h = 1, 2, \ldots, 28$ are presented in the last three columns. The analysis of forecast accuracy over different horizons is fundamental to assist the management of call centers. In fact, while forecasts at monthly horizon are needed for hiring new agents, forecasts at weekly and daily horizon are used for the scheduling of the available pool of agents (Akşin et al., 2007).

Focusing on single models, for both univariate and multivariate losses the SARMAX–GARCH is always among the best–performing specifications. Looking at the symmetric multivariate loss in column 10 of Table 4, we see that this result holds also when including the combined forecasts in the ranking. As for the combining methods, the ABMA based on the AIC seems to be the best available option in the majority of cases. Notice that the best ABMA combinations are those based on groups of models that include the SARMAX forecasts with and without GARCH equation and the ARMAX-GARCH forecasts, namely those individual forecasts associated with some of the lowest individual and system losses.

When analyzing asymmetric losses, the ranking of models changes according to the incidence of over– and under–forecasting. Nevertheless, we can confirm most of the results just highlighted for the symmetric case. Interestingly, in the multivariate case, the MEM becomes the best option when under–forecasting is more penalized than over–forecasting ($\phi = 0.58$).

Focusing on combination schemes, "minimum forecasts" based on the groups of models $\mathcal{G}_3$ and $\mathcal{G}_4$ yield the lowest average losses when $\phi = 0.42$. On the contrary, when under–forecasting in costlier than over–forecasting ($\phi = 0.58$), these forecasts are not optimal anymore. In this case, either the ABMA–AIC combining methods or the "maximum forecasts" lead to the lowest average losses. Moreover, we can observe that neither count data models, nor the Spline–SARX model seem to be valuable options for obtaining accurate point forecasts.

Lastly, we find that second–moment modeling is important when forecasting call arrivals. In fact, when a GARCH component is added to ARMAX and SARMAX models, their ranking improves in most cases. This result suggests (see Section 4.4) that these simple models could also be useful in the literature dealing with density forecasts (see e.g. Taylor, 2012).

From the standing point of a practitioner, results in Table 4 also suggest that, independently of the shape of the loss function, investing in model building or outsourcing the forecasting exercise could be worth its cost. In fact, the benchmark SRW model is always outperformed by other relatively more general specifications.[10]

Since the number of forecasts under consideration is quite large, these conclusions might

[10]This fact is supported also by the Diebold and Mariano (1995) test. These results are available from the authors upon request.

Table 4: Ranking of models and combined forecasts

| | $h=1$ | | | $h=7$ | | | $h=28$ | | | $h=1,...,28$ | | |
|---|---|---|---|---|---|---|---|---|---|---|---|---|
| | $\phi=0.42$ | $\phi=0.5$ | $\phi=0.58$ | $\phi=0.42$ | $\phi=0.5$ | $\phi=0.58$ | $\phi=0.42$ | $\phi=0.5$ | $\phi=0.58$ | $\phi=0.42$ | $\phi=0.5$ | $\phi=0.58$ |
| $\mathcal{M}_0$ | 44 (13) | 44 (13) | 44 (13) | 30 (7) | 32 (7) | 31 (7) | 41 (10) | 43 (12) | 43 (12) | 29 (7) | 43 (12) | 43 (12) |
| $\mathcal{M}_1$ | 18 (4) | 21 (5) | 22 (6) | 19 (4) | 19 (4) | 18 (4) | 27 (5) | 26 (5) | 23 (4) | 26 (5) | 23 (5) | 22 (6) |
| $\mathcal{M}_2$ | 21 (5) | 18 (4) | 17 (**3**) | 14 (**3**) | 14 (**3**) | 13 (**3**) | 16 (**3**) | 16 (**3**) | 15 (**3**) | 15 (**3**) | 14 (**3**) | 28 (7) |
| $\mathcal{M}_3$ | 26 (8) | 24 (7) | 24 (7) | 28 (6) | 27 (6) | 28 (6) | 20 (4) | 23 (4) | 24 (5) | 18 (4) | 22 (4) | 37 (10) |
| $\mathcal{M}_4$ | 13 (**3**) | 16 (**3**) | 21 (5) | 7 (**2**) | 5 (**2**) | 5 (**2**) | 6 (**2**) | 6 (**2**) | 4 (**2**) | 8 (**2**) | 6 (**2**) | 31 (8) |
| $\mathcal{M}_5$ | 9 (**2**) | 9 (**2**) | 9 (**1**) | 4 (**1**) | **2** (**1**) | **2** (**1**) | 5 (**1**) | **3** (**1**) | **3** (**1**) | 4 (**1**) | **1** (**1**) | 34 (9) |
| $\mathcal{M}_6$ | 23 (6) | 22 (6) | 20 (4) | 21 (5) | 20 (5) | 21 (5) | 28 (6) | 27 (6) | 27 (6) | 28 (6) | 26 (6) | 8 (**3**) |
| $\mathcal{M}_7$ | 5 (**1**) | 8 (**1**) | 16 (**2**) | 34 (9) | 43 (12) | 43 (12) | 45 (14) | 45 (14) | 45 (14) | 44 (13) | 45 (14) | 45 (14) |
| $\mathcal{M}_8$ | 41 (12) | 41 (12) | 40 (12) | 41 (12) | 40 (11) | 39 (11) | 38 (9) | 37 (9) | 37 (9) | 40 (11) | 37 (9) | 6 (**2**) |
| $\mathcal{M}_9$ | 37 (10) | 37 (10) | 37 (10) | 38 (11) | 37 (10) | 36 (10) | 34 (8) | 33 (8) | 34 (8) | 37 (10) | 34 (8) | 14 (4) |
| $\mathcal{M}_{10}$ | 39 (11) | 39 (11) | 39 (11) | 36 (10) | 35 (9) | 34 (9) | 32 (7) | 31 (7) | 32 (7) | 35 (9) | 32 (7) | 16 (5) |
| $\mathcal{M}_{11}$ | 45 (14) | 46 (14) | 46 (14) | 44 (13) | 44 (13) | 44 (13) | 43 (12) | 41 (10) | 40 (10) | 43 (12) | 42 (11) | **1** (**1**) |
| $\mathcal{M}_{12}$ | 29 (9) | 30 (9) | 32 (9) | 45 (14) | 45 (14) | 45 (14) | 44 (13) | 44 (13) | 44 (13) | 45 (14) | 44 (13) | 44 (13) |
| $\mathcal{M}_{13}$ | 24 (7) | 27 (8) | 30 (8) | 33 (8) | 33 (8) | 32 (8) | 42 (11) | 42 (11) | 42 (11) | 30 (8) | 41 (10) | 42 (11) |
| $Avg.\mathcal{G}_1$ | 16 [13] | 15 [13] | 12 [11] | 23 [18] | 21 [16] | 20 [16] | 13 [11] | 12 [10] | 12 [10] | 25 [21] | 18 [15] | 9 [6] |
| $Avg.\mathcal{G}_2$ | 17 [14] | 17 [14] | 14 [13] | 16 [13] | 16 [13] | 12 [10] | 19 [16] | 18 [15] | 17 [14] | 17 [14] | 17 [14] | 26 [20] |
| $Avg.\mathcal{G}_3$ | 6 [5] | 5 [5] | **3** [**3**] | 13 [11] | 12 [10] | 10 [8] | 12 [10] | 11 [9] | 11 [9] | 12 [10] | 11 [9] | 19 [14] |
| $Avg.\mathcal{G}_4$ | 12 [10] | 12 [10] | 8 [8] | 11 [9] | 10 [8] | 9 [7] | 8 [6] | 9 [7] | 7 [5] | 11 [9] | 9 [7] | 25 [19] |
| $Avg.\mathcal{G}_5$ | 40 [29] | 40 [29] | 35 [26] | 40 [29] | 39 [29] | 38 [28] | 36 [28] | 36 [28] | 36 [28] | 39 [29] | 36 [28] | 11 [8] |
| $Tr.Avg.\mathcal{G}_1$ | 8 [7] | 6 [6] | 6 [6] | 22 [17] | 23 [18] | 22 [17] | 11 [9] | 10 [8] | 10 [8] | 20 [16] | 16 [13] | 12 [9] |
| $Tr.Avg.\mathcal{G}_2$ | 25 [18] | 28 [20] | 29 [22] | 20 [16] | 22 [17] | 23 [18] | 21 [17] | 20 [17] | 19 [16] | 21 [17] | 21 [18] | 27 [21] |
| $Tr.Avg.\mathcal{G}_3$ | 15 [12] | 14 [12] | 15 [14] | 12 [10] | 11 [9] | 8 [6] | 14 [12] | 13 [11] | 13 [11] | 13 [11] | 13 [11] | 20 [15] |
| $Tr.Avg.\mathcal{G}_4$ | 19 [15] | 23 [17] | 23 [17] | 10 [8] | 9 [7] | 7 [5] | 7 [5] | 7 [5] | 5 [**3**] | 10 [8] | 8 [6] | 24 [18] |
| $Tr.Avg.\mathcal{G}_5$ | 47 [33] | 47 [33] | 47 [33] | 47 [33] | 47 [33] | 47 [33] | 47 [33] | 47 [33] | 47 [33] | 46 [32] | 47 [33] | 47 [33] |

*Notes*: continued on next page

Table 4 (*continued*)

Ranking of models and combined forecasts

| | $h=1$ | | | $h=7$ | | | $h=28$ | | | $h=1,...,28$ | | |
|---|---|---|---|---|---|---|---|---|---|---|---|---|
| | $\phi=0.42$ | $\phi=0.5$ | $\phi=0.58$ | $\phi=0.42$ | $\phi=0.5$ | $\phi=0.58$ | $\phi=0.42$ | $\phi=0.5$ | $\phi=0.58$ | $\phi=0.42$ | $\phi=0.5$ | $\phi=0.58$ |
| $Med.\mathcal{G}_1$ | 4 [4] | 4 [4] | **1** [**1**] | 18 [15] | 17 [14] | 11 [9] | 23 [19] | 19 [16] | 18 [15] | 24 [20] | 19 [16] | 10 [7] |
| $Med.\mathcal{G}_2$ | 20 [16] | 20 [16] | 19 [16] | 17 [14] | 18 [15] | 15 [12] | 21 [17] | 20 [17] | 19 [16] | 19 [15] | 20 [17] | 23 [17] |
| $Med.\mathcal{G}_3$ | 7 [6] | 7 [7] | 7 [7] | 9 [7] | 8 [6] | 6 [4] | 15 [13] | 14 [12] | 14 [12] | 14 [12] | 12 [10] | 18 [13] |
| $Med.\mathcal{G}_4$ | 14 [11] | 13 [11] | 13 [12] | 8 [6] | 4 [**3**] | 4 [**3**] | 9 [7] | 8 [6] | 6 [4] | 9 [7] | 7 [5] | 21 [16] |
| $Med.\mathcal{G}_5$ | 36 [27] | 36 [27] | 36 [27] | 38 [27] | 37 [27] | 36 [26] | 34 [26] | 33 [25] | 34 [26] | 36 [27] | 33 [26] | 13 [10] |
| $Min\mathcal{G}_1$ | 31 [22] | 34 [25] | 43 [31] | 24 [19] | 31 [25] | 41 [30] | 18 [15] | 40 [31] | 41 [31] | **1** [**1**] | 40 [31] | 46 [32] |
| $Min\mathcal{G}_2$ | **3** [**3**] | **3** [**3**] | 5 [5] | 6 [5] | 13 [11] | 19 [15] | 10 [8] | 15 [13] | 22 [19] | 7 [6] | 10 [8] | 38 [28] |
| $Min\mathcal{G}_3$ | **1** [**1**] | **1** [**1**] | **2** [**2**] | **1** [**1**] | 6 [4] | 17 [14] | **2** [**2**] | 5 [4] | 9 [7] | **2** [**2**] | 4 [**3**] | 40 [30] |
| $Min\mathcal{G}_4$ | **2** [**2**] | **2** [**2**] | 4 [4] | **2** [**2**] | 7 [5] | 16 [13] | **1** [**1**] | 4 [**3**] | 8 [6] | **3** [**3**] | 5 [4] | 39 [29] |
| $Min\mathcal{G}_5$ | 35 [26] | 35 [26] | 34 [25] | 35 [26] | 34 [26] | 33 [25] | 30 [24] | 29 [23] | 30 [24] | 34 [26] | 31 [25] | 17 [12] |
| $Max\mathcal{G}_1$ | 46 [32] | 45 [32] | 45 [32] | 46 [32] | 46 [32] | 46 [32] | 46 [32] | 46 [32] | 46 [32] | 47 [33] | 46 [32] | 41 [31] |
| $Max\mathcal{G}_2$ | 32 [23] | 31 [22] | 28 [21] | 29 [23] | 28 [22] | 27 [22] | 29 [23] | 28 [22] | 28 [22] | 31 [23] | 28 [22] | 4 [**3**] |
| $Max\mathcal{G}_3$ | 34 [25] | 33 [24] | 33 [24] | 32 [25] | 30 [24] | 30 [24] | 37 [29] | 35 [27] | 31 [25] | 33 [25] | 30 [24] | **2** [**1**] |
| $Max\mathcal{G}_4$ | 33 [24] | 32 [23] | 31 [23] | 31 [24] | 29 [23] | 29 [23] | 31 [25] | 30 [24] | 29 [23] | 32 [24] | 29 [23] | **3** [**2**] |
| $Max\mathcal{G}_5$ | 43 [31] | 43 [31] | 42 [30] | 43 [31] | 42 [31] | 42 [31] | 40 [31] | 39 [30] | 39 [30] | 42 [31] | 39 [30] | 5 [4] |
| $SIC\mathcal{G}_2$ | 28 [20] | 26 [19] | 26 [19] | 27 [22] | 26 [21] | 26 [21] | 25 [21] | 25 [21] | 26 [21] | 23 [19] | 25 [20] | 36 [27] |
| $SIC\mathcal{G}_3$ | 30 [21] | 29 [21] | 27 [20] | 25 [20] | 24 [19] | 24 [19] | 26 [22] | 22 [19] | 21 [18] | 27 [22] | 27 [21] | 30 [23] |
| $SIC\mathcal{G}_4$ | 27 [19] | 25 [18] | 25 [18] | 26 [21] | 25 [20] | 25 [20] | 24 [20] | 24 [20] | 25 [20] | 22 [18] | 24 [19] | 35 [26] |
| $SIC\mathcal{G}_5$ | 41 [30] | 41 [30] | 40 [29] | 41 [30] | 40 [30] | 39 [29] | 38 [30] | 37 [29] | 37 [29] | 40 [30] | 37 [29] | 6 [5] |
| $AIC\mathcal{G}_2$ | 21 [17] | 18 [15] | 17 [15] | 14 [12] | 14 [12] | 13 [11] | 16 [14] | 16 [14] | 15 [13] | 15 [13] | 14 [12] | 28 [22] |
| $AIC\mathcal{G}_3$ | 10 [8] | 10 [8] | 10 [9] | 4 [**3**] | **2** [**1**] | **2** [**1**] | **3** [**3**] | **1** [**1**] | **1** [**1**] | 5 [4] | **2** [**1**] | 32 [24] |
| $AIC\mathcal{G}_4$ | 10 [8] | 10 [8] | 10 [9] | 4 [**3**] | **2** [**1**] | **2** [**1**] | **3** [**3**] | **1** [**1**] | **1** [**1**] | 5 [4] | **2** [**1**] | 32 [24] |
| $AIC\mathcal{G}_5$ | 37 [28] | 37 [28] | 37 [28] | 38 [27] | 37 [27] | 36 [26] | 34 [26] | 33 [25] | 34 [26] | 37 [28] | 34 [27] | 14 [11] |

*Notes*: the first column uses the following shorthand notation: "Avg." = Average, "Tr. Avg." = Trimmed Average, "Med." = Median, "SIC" = ABMA using SIC and "AIC" = ABMA using AIC; entries represent the ranking of models and forecast combinations based on the statistics $\sqrt{P^{-1}\sum_{t=1}^{P}\mathcal{L}_t}$, where $\mathcal{L}_{\sqcup}$ is the generalized loss function in Eq. (1) and $P = 351$; statistics in columns 2-10 are based on the univariate loss for forecast horizons, $h = 1, 7, 28$, while statistics in the last three columns are based on the multivariate loss for $h = 1, ..., 28$; the shape of the loss function is determined by $\rho = 2$ and the asymmetry coefficient: $\phi = 0.42, 0.50, 0.58$ ($\tau = -0.16, 0.00, 0.16$); these values guarantee that the multivariate loss is always non-negative (see Komunjer and Owyang (2012) for details); entries outside brackets represent the overall ranking of the forecast; entries in round brackets represent the ranking among models, while entries in square brackets denote the ranking among combining methods; models, $\mathcal{M}_m$, groups of models, $\mathcal{G}_i$, and combining methods are described in Tables 1 and 2; entries in bold denote models with the lowest loss.

Table 5: Reality Check test

| | $h=1$ | | | $h=7$ | | | $h=28$ | | |
|---|---|---|---|---|---|---|---|---|---|
| | $\phi$ | | | $\phi$ | | | $\phi$ | | |
| | 0.42 | 0.50 | 0.58 | 0.42 | 0.50 | 0.58 | 0.42 | 0.50 | 0.58 |
| $\mathcal{M}_0$ | – | – | – | $\star$ | – | – | $\star\star$ | $\star\star$ | $\star\star$ |
| $\mathcal{M}_1$ | – | – | $\star$ | – | – | – | – | – | – |
| $\mathcal{M}_2$ | – | – | $\star$ | $\star$ | $\star$ | $\star$ | – | – | $\star$ |
| $\mathcal{M}_3$ | – | – | $\star$ | – | – | – | – | $\star$ | $\star$ |
| $\mathcal{M}_4$ | – | – | $\star\star$ | $\star\star$ | $\star\star$ | $\star\star$ | – | $\star$ | $\star\star$ |
| $\mathcal{M}_5$ | – | $\star\star$ | $\star\star$ | $\star\star$ | $\star\star$ | $\star\star$ | $\star\star$ | $\star\star$ | $\star\star$ |
| $\mathcal{M}_6$ | $\star\star$ | $\star\star$ | $\star\star$ | $\star$ | $\star\star$ | $\star\star$ | – | – | $\star$ |
| $\mathcal{M}_7$ | $\star$ | $\star\star$ | $\star\star$ | – | – | – | $\star$ | $\star$ | $\star$ |
| $\mathcal{M}_8$ | – | – | – | – | – | – | – | – | – |
| $\mathcal{M}_9$ | – | – | – | – | – | – | – | – | – |
| $\mathcal{M}_{10}$ | – | – | – | – | – | – | – | – | – |
| $\mathcal{M}_{11}$ | – | – | – | – | – | – | – | – | – |
| $\mathcal{M}_{12}$ | – | – | – | – | – | – | – | – | – |
| $\mathcal{M}_{13}$ | – | – | – | – | – | – | $\star\star$ | $\star\star$ | $\star\star$ |
| Avg. $\mathcal{G}_1$ | – | – | $\star$ | – | $\star$ | $\star\star$ | – | $\star$ | $\star\star$ |
| Avg. $\mathcal{G}_2$ | – | $\star\star$ | $\star\star$ | – | – | – | – | – | – |
| Avg. $\mathcal{G}_3$ | $\star\star$ | $\star\star$ | $\star\star$ | – | – | $\star$ | – | – | – |
| Avg. $\mathcal{G}_4$ | – | $\star\star$ | $\star\star$ | $\star$ | $\star$ | $\star\star$ | $\star\star$ | $\star\star$ | $\star\star$ |
| Avg. $\mathcal{G}_5$ | – | – | – | – | – | – | – | – | $\star$ |
| Tr. Avg. $\mathcal{G}_1$ | – | $\star$ | $\star$ | $\star$ | – | $\star$ | – | $\star$ | $\star\star$ |
| Tr. Avg. $\mathcal{G}_2$ | – | $\star$ | $\star\star$ | – | $\star$ | $\star$ | – | – | – |
| Tr. Avg. $\mathcal{G}_3$ | – | $\star$ | $\star\star$ | $\star$ | $\star$ | $\star\star$ | – | – | $\star\star$ |
| Tr. Avg. $\mathcal{G}_4$ | – | – | – | $\star$ | – | – | $\star\star$ | $\star\star$ | $\star\star$ |
| Tr. Avg. $\mathcal{G}_5$ | – | – | – | – | – | – | – | – | – |
| Med. $\mathcal{G}_1$ | $\star$ | $\star\star$ | $\star\star$ | – | $\star\star$ | $\star\star$ | – | – | – |
| Med. $\mathcal{G}_2$ | – | – | – | – | – | $\star$ | – | – | – |
| Med. $\mathcal{G}_3$ | – | $\star\star$ | $\star\star$ | $\star\star$ | $\star\star$ | $\star\star$ | – | – | $\star\star$ |
| Med. $\mathcal{G}_4$ | – | $\star$ | $\star\star$ | $\star\star$ | $\star\star$ | $\star\star$ | $\star\star$ | $\star\star$ | $\star\star$ |
| Med. $\mathcal{G}_5$ | – | – | – | – | – | – | – | – | – |
| Min $\mathcal{G}_1$ | $\star\star$ | $\star\star$ | $\star\star$ | – | – | – | $\star\star$ | $\star\star$ | $\star\star$ |
| Min $\mathcal{G}_2$ | $\star\star$ | $\star\star$ | $\star\star$ | $\star$ | $\star$ | $\star\star$ | $\star\star$ | $\star\star$ | $\star\star$ |
| Min $\mathcal{G}_3$ | $\star\star$ | $\star\star$ | $\star\star$ | $\star\star$ | $\star\star$ | $\star\star$ | $\star\star$ | $\star\star$ | $\star\star$ |
| Min $\mathcal{G}_4$ | $\star\star$ | $\star\star$ | $\star\star$ | $\star\star$ | $\star\star$ | $\star\star$ | $\star\star$ | $\star\star$ | $\star\star$ |
| Min $\mathcal{G}_5$ | – | – | – | – | – | – | – | – | – |
| Max $\mathcal{G}_1$ | – | – | – | – | – | – | – | – | – |
| Max $\mathcal{G}_2$ | – | – | – | – | – | – | – | – | $\star$ |
| Max $\mathcal{G}_3$ | – | – | – | – | – | – | – | – | – |
| Max $\mathcal{G}_4$ | – | – | – | – | – | – | – | – | – |
| Max $\mathcal{G}_5$ | – | – | – | – | – | – | – | – | – |
| SIC $\mathcal{G}_2$ | – | – | – | – | $\star$ | $\star$ | – | – | $\star$ |
| SIC $\mathcal{G}_3$ | – | – | – | – | – | – | – | $\star$ | $\star\star$ |
| SIC $\mathcal{G}_4$ | – | – | – | – | – | $\star$ | – | – | $\star$ |
| SIC $\mathcal{G}_5$ | – | – | – | – | – | – | – | – | – |
| AIC $\mathcal{G}_2$ | – | – | $\star$ | $\star$ | $\star\star$ | $\star\star$ | – | $\star$ | $\star$ |
| AIC $\mathcal{G}_3$ | – | $\star\star$ | $\star\star$ | $\star\star$ | $\star\star$ | $\star\star$ | $\star\star$ | $\star\star$ | $\star\star$ |
| AIC $\mathcal{G}_4$ | – | $\star\star$ | $\star\star$ | $\star\star$ | $\star\star$ | $\star\star$ | $\star\star$ | $\star\star$ | $\star\star$ |
| AIC $\mathcal{G}_5$ | – | – | – | – | – | – | – | – | – |

*Notes*: the table presents results of the Reality Check test of White (2000), as modified by Hansen (2005); the benchmark model is indicated in the first column, where the following shorthand notation is used: "Avg." = Average, "Tr. Avg." = Trimmed Average, "Med." = Median, "SIC" = ABMA using SIC and "AIC" = ABMA using AIC; the test is implemented using the stationary (block) bootstrap of Politis and Romano (1994); the number of bootstrap repetitions is equal to 999, the block length equals 29 days; a p-value lower than 0.05 indicates rejection of the null hypothesis that the benchmark performs as well as the best alternative model; "–" denotes a p-value $< 0.05$, "$\star$" denotes $0.05 \leq$ p-value $< 0.1$, "$\star\star$" denotes a p-value $\geq 0.1$.

be subject to data snooping effects. Both the Reality Check test (RCT) of White (2000) and the Model Confidence Set (MCS) of Hansen et al. (2011) are designed to deal with data snooping. The difference between the two procedures is that the former requires a benchmark model, while the latter does not. Results of the RCT test of the null hypothesis that the benchmark performs as well as the best alternative model are shown in Table 5. Our findings are based on the consistent $p$–values of Hansen (2005), who has shown that the

Table 6: Are models within the MCS?

| | $h=1$ | | | $h=7$ | | | $h=28$ | | |
|---|---|---|---|---|---|---|---|---|---|
| | $\phi=0.42$ | $\phi=0.5$ | $\phi=0.58$ | $\phi=0.42$ | $\phi=0.5$ | $\phi=0.58$ | $\phi=0.42$ | $\phi=0.5$ | $\phi=0.58$ |
| $\mathcal{M}_0$ | – | – | – | – | – | – | – | – | – |
| $\mathcal{M}_1$ | – | – | – | – | – | – | – | – | – |
| $\mathcal{M}_2$ | – | – | – | – | – | – | – | – | – |
| $\mathcal{M}_3$ | – | – | – | – | – | – | – | – | – |
| $\mathcal{M}_4$ | – | ✓ | – | – | – | – | – | – | – |
| $\mathcal{M}_5$ | – | ✓ | ✓ | – | – | – | – | – | – |
| $\mathcal{M}_6$ | – | ✓ | ✓ | – | – | – | – | – | – |
| $\mathcal{M}_7$ | ✓ | ✓ | ✓ | – | – | – | – | – | – |
| $\mathcal{M}_8$ | – | – | – | – | – | – | – | – | – |
| $\mathcal{M}_9$ | – | – | – | – | – | – | – | – | – |
| $\mathcal{M}_{10}$ | – | – | – | – | – | – | – | – | – |
| $\mathcal{M}_{11}$ | – | – | – | – | – | – | – | – | – |
| $\mathcal{M}_{12}$ | – | – | – | – | – | – | – | – | – |
| $\mathcal{M}_{13}$ | – | – | – | – | – | – | – | – | – |
| Avg. $\mathcal{G}_1$ | – | – | ✓ | – | – | – | – | – | – |
| Avg. $\mathcal{G}_2$ | – | ✓ | ✓ | – | – | – | – | – | – |
| Avg. $\mathcal{G}_3$ | – | ✓ | ✓ | – | – | – | – | – | – |
| Avg. $\mathcal{G}_4$ | – | ✓ | ✓ | – | – | – | – | – | – |
| Avg. $\mathcal{G}_5$ | – | – | – | – | – | – | – | – | – |
| Tr. Avg. $\mathcal{G}_1$ | – | ✓ | ✓ | – | – | – | – | – | – |
| Tr. Avg. $\mathcal{G}_2$ | – | – | – | – | – | – | – | – | – |
| Tr. Avg. $\mathcal{G}_3$ | – | – | – | – | – | – | – | – | – |
| Tr. Avg. $\mathcal{G}_4$ | – | – | – | – | – | – | – | – | – |
| Tr. Avg. $\mathcal{G}_5$ | – | – | – | – | – | – | – | – | – |
| Med. $\mathcal{G}_1$ | – | ✓ | ✓ | – | – | – | – | – | – |
| Med. $\mathcal{G}_2$ | – | – | – | – | – | – | – | – | – |
| Med. $\mathcal{G}_3$ | – | ✓ | ✓ | – | – | – | – | – | – |
| Med. $\mathcal{G}_4$ | – | – | – | – | – | – | – | – | – |
| Med. $\mathcal{G}_5$ | – | – | – | – | – | – | – | – | – |
| Min $\mathcal{G}_1$ | – | – | – | – | – | – | – | – | – |
| Min $\mathcal{G}_2$ | ✓ | ✓ | ✓ | – | – | – | – | – | – |
| Min $\mathcal{G}_3$ | ✓ | ✓ | ✓ | ✓ | – | – | ✓ | – | – |
| Min $\mathcal{G}_4$ | ✓ | ✓ | ✓ | ✓ | – | – | ✓ | – | – |
| Min $\mathcal{G}_5$ | – | – | – | – | – | – | – | – | – |
| Max $\mathcal{G}_1$ | – | – | – | – | – | – | – | – | – |
| Max $\mathcal{G}_2$ | – | – | – | – | – | – | – | – | – |
| Max $\mathcal{G}_3$ | – | – | – | – | – | – | – | – | – |
| Max $\mathcal{G}_4$ | – | – | – | – | – | – | – | – | – |
| Max $\mathcal{G}_5$ | – | – | – | – | – | – | – | – | – |
| SIC $\mathcal{G}_2$ | – | – | – | – | – | – | – | – | – |
| SIC $\mathcal{G}_3$ | – | – | – | – | – | – | – | – | – |
| SIC $\mathcal{G}_4$ | – | – | – | – | – | – | – | – | – |
| SIC $\mathcal{G}_5$ | – | – | – | – | – | – | – | – | – |
| AIC $\mathcal{G}_2$ | – | – | – | – | – | – | – | – | – |
| AIC $\mathcal{G}_3$ | – | – | – | – | – | – | – | – | – |
| AIC $\mathcal{G}_4$ | – | ✓ | ✓ | ✓ | ✓ | ✓ | ✓ | ✓ | ✓ |
| AIC $\mathcal{G}_5$ | – | – | – | – | – | – | – | – | – |

*Notes*: the table presents results of the Model Confidence Set (MCS) procedure of Hansen et al. (2011), implemented using the stationary (block) bootstrap of Politis and Romano (1994); the number of bootstrap repetitions is equal to 999, the block length equals 29 days; "–" indicates that the model is not in the MCS at the 90% confidence level, while "✓" indicates that the model belongs to the MCS; the first column uses the following shorthand notation: "Avg." = Average, "Tr. Avg." = Trimmed Average, "Med." = Median, "SIC" = ABMA using SIC and "AIC" = ABMA using AIC.

original procedure has low power when a poor performing forecast enters the set of alternative models. Using the SARMAX–GARCH model, or the ABMA–AIC combined forecasts (based on the models belonging to group $\mathcal{G}_3$ or group $\mathcal{G}_4$), we reject the null hypothesis only once.

The MCS test is used to compare the forecast accuracy of models without selecting a benchmark model and yields a set of specifications that contains the best forecast with a prespecified asymptotic probability. As it can be seen from Table 6, this test is more selective than the RCT. Considering one day ahead forecasts and under MSE loss, the MCS at the 90% confidence level contains only four individual models: SARMAX, SARMAX–GARCH, PAR and Airline. When over–forecasting is more penalized than under–prediction (i.e. $\phi = 0.42$), the only model entering the MCS is the Airline. On the contrary, when positive forecast errors are more heavily weighted than negative forecast errors (i.e. $\phi = 0.58$), the SARMAX–GARCH and the PAR are also in the MCS. When the forecast horizon is one week, or one month, only the ABMA–AIC combined forecasts based on model group $\mathcal{G}_4$ are always in the MCS.

Therefore, when the loss function is parametrized so as to penalize under–prediction more heavily than over–prediction, the best performing model and combination method are the SARMAX–GARCH and the ABMA–AIC based on model group $\mathcal{G}_4$.

## 3.2 Monetary measures

We now focus on the task of selecting the best forecasting approach, given a forecast horizon of one day and assuming that the manager of the call center is more adverse to under–staffing than to over–staffing. We thus define the set of alternative forecasts so as to include all individual models and the combined forecast obtained with ABMA–AIC applied to group $\mathcal{G}_4$.

The economic evaluation of forecasts based on the willingness to pay, $\delta_i$, and the incremental value of information, $\Delta V_i$, is presented in Table 7. Although both economic measures of performance decrease as the absolute risk aversion increases, the manager's willingness to pay seems to be less responsive to such a change than the incremental value of information.

The second column of Table 7 shows the percentage change in the root MSE distance for comparison. An entry below 100 indicates that the model $\mathcal{M}_i$ outperforms the benchmark. All measures suggest that the worst performing model is MEM, while the best models are the SARMAX–GARCH and the ABMA–AIC combined forecasts. The manager is willing to pay up to 1687 Euro in order to use these models instead of the benchmark. The model characterized by the minimum (positive) willingness to pay, 912 Euro, is the Poisson count data specification.

The ranking based on the incremental value of information is consistent with that based on the willingness to pay. On the contrary, being symmetric about zero forecast errors, the ranking based on the MSE ranking is quite different. For instance, the Airline model would be preferred to the SARMAX–GARCH model, which is the best option when the loss function is consistent with the manager's compensation scheme.

Overall, both economic and statistical evaluation of models indicate the SARMAX–GARCH model and the ABMA–AIC combining method based on model group $\mathcal{G}_4$ as the best options for the manager. However, using only statistical methods, we cannot clearly identify which of these options is best. On the contrary, given that the monetary value of the two forecasts and the manager's willingness to pay are very similar, we can conclude that the SARMAX–GARCH model is to be preferred to the ABMA–AIC forecast combination method.

Table 7: Economic evaluation of models

| | | $\delta_i$ (Euro) | | | $\Delta V_i$ (Euro) | | |
|---|---|---|---|---|---|---|---|
| | ΔRMSE (%) | $\lambda = 0.0002$ | $\lambda = 0.0003$ | $\lambda = 0.0005$ | $\lambda = 0.0002$ | $\lambda = 0.0003$ | $\lambda = 0.0005$ |
| $\mathcal{M}_1$ | 58.40 | 1377.40 | 1377.40 | 1377.30 | 401.66 | 114.03 | 11.96 |
| $\mathcal{M}_2$ | 58.50 | 1544.90 | 1544.80 | 1544.60 | 445.01 | 125.19 | 12.97 |
| $\mathcal{M}_3$ | 62.27 | 1342.40 | 1342.40 | 1342.30 | 392.47 | 111.63 | 11.74 |
| $\mathcal{M}_4$ | 57.06 | 1413.70 | 1413.70 | 1413.70 | 411.15 | 116.49 | 12.19 |
| $\mathcal{M}_5$ | 56.85 | 1687.40 | 1687.30 | 1687.20 | 481.04 | 134.29 | 13.77 |
| $\mathcal{M}_6$ | 59.02 | 1433.70 | 1433.60 | 1433.50 | 416.35 | 117.84 | 12.31 |
| $\mathcal{M}_7$ | 55.86 | 1517.40 | 1517.40 | 1517.30 | 437.98 | 123.40 | 12.81 |
| $\mathcal{M}_8$ | 76.36 | 912.39 | 912.30 | 912.16 | 275.33 | 80.22 | 8.73 |
| $\mathcal{M}_9$ | 76.10 | 988.59 | 988.47 | 988.28 | 296.65 | 86.05 | 9.30 |
| $\mathcal{M}_{10}$ | 76.10 | 988.59 | 988.47 | 988.28 | 296.65 | 86.05 | 9.30 |
| $\mathcal{M}_{11}$ | 123.66 | -933.48 | -933.27 | -932.94 | -324.07 | -105.60 | -13.59 |
| $\mathcal{M}_{12}$ | 62.72 | 1031.10 | 1031.10 | 1030.90 | 308.44 | 89.26 | 9.62 |
| $\mathcal{M}_{13}$ | 61.19 | 1217.50 | 1217.40 | 1217.40 | 359.21 | 102.88 | 10.93 |
| AIC $\mathcal{G}_4$ | 56.85 | 1687.40 | 1687.30 | 1687.20 | 481.04 | 134.29 | 13.77 |

*Notes*: economic evaluation of one day ahead forecasts; ΔRMSE=100×(RMSE$_i$/RMSE$_0$), where the RMSE corresponds to the flexible loss distance for $\rho = 2$ and $\phi = 0.5$; the incremental value of information is $\Delta V_i = V_i - V_0$, where $V_i$ is the value of information from model $i$; the willingness to pay for model $i$ is $\delta = CE_i - CE_0$, where $CE_i$ is the certainty equivalent from model $i$; $\lambda$ is the coefficient of risk aversion.

Actually, given that the latter method involves more than one model and that the specification of models has to be periodically revised, the ABMA–AIC forecast combination method will have higher maintenance costs than the SARMAX–GARCH model. Moreover, if the model is run by an employee of the call center and not by the professional forecaster, we see the "ease–of–use" as a critical factor for the choice of the best forecast.

All in all, we have shown that simple measures of performance expressed in monetary terms are easy to construct and are more flexible than the often used symmetric loss functions. This flexibility allows the forecasts' user and the adviser to judge the predictive performance of models with the same metric. Moreover, being expressed in monetary terms, we believe that these measures are more interesting for practitioners than traditional statistical distances. Finally, from the perspective of the professional forecaster, we see the results in this section as complementary to those based on flexible loss functions. From the standpoint of forecast users, we believe that the economic measures are to be preferred, as they are more closely linked to their profit maximizing behavior.

# 4 Robustness checks

## 4.1 Rolling estimation window

Although the recursive (or expanding) scheme has the advantage of using more observations than its rolling (or moving) window counterpart, the latter is more robust to the presence of structural breaks, while it is based on an estimation sample size which is arbitrarily chosen. The fixed forecasting scheme is another option, which not only does involve an arbitrary choice of the sample size, but also it is less robust to the presence of structural breaks in the variance and in the mean of the series. Its main strength is the fact that the model is estimated only once. However, this aspect does not justify its use in our setting, given that all models we have proposed can be easily estimated and forecasts and simulations are based on the most recent sample of observations.

For these reasons, we have decided to produce new forecasts for the Italian call center with a rolling scheme and a sample size $R = 371$. This sample size matches that of the first estimation sample used with the recursive scheme, and it allows to rely on the same evaluation sample. Results in Table 8 show that SARMAX specifications, with or without the GARCH component, remain among the preferred options at all forecast horizons, irrespective of the shape of the loss function.

Table 8: Ranking of models for the Italian call center (rolling window estimation)

| | (a) $h=1$ | | | (b) $h=7$ | | |
|---|---|---|---|---|---|---|
| | $\phi=0.42$ | $\phi=0.5$ | $\phi=0.58$ | $\phi=0.42$ | $\phi=0.5$ | $\phi=0.58$ |
| $\mathcal{M}_0$ | 14 | 14 | 14 | 14 | 14 | 14 |
| $\mathcal{M}_1$ | **3** | **3** | 4 | 4 | 4 | 4 |
| $\mathcal{M}_2$ | 5 | 4 | **1** | 5 | 5 | 5 |
| $\mathcal{M}_3$ | 7 | 7 | 9 | 10 | 12 | 13 |
| $\mathcal{M}_4$ | **2** | **1** | **3** | **1** | **1** | **1** |
| $\mathcal{M}_5$ | 4 | **2** | **2** | **3** | **3** | **2** |
| $\mathcal{M}_6$ | 9 | 8 | 7 | **2** | **2** | **3** |
| $\mathcal{M}_7$ | 6 | 6 | 6 | 13 | 13 | 12 |
| $\mathcal{M}_8$ | 10 | 10 | 10 | 9 | 9 | 8 |
| $\mathcal{M}_9$ | 11 | 11 | 11 | 8 | 8 | 7 |
| $\mathcal{M}_{10}$ | 12 | 12 | 12 | 7 | 7 | 6 |
| $\mathcal{M}_{11}$ | 13 | 13 | 13 | 11 | 10 | 9 |
| $\mathcal{M}_{12}$ | **1** | 5 | 5 | 6 | 6 | 10 |
| $\mathcal{M}_{13}$ | 8 | 9 | 8 | 12 | 11 | 11 |
| | (c) $h=28$ | | | (d) $h=1,...,28$ | | |
| | $\phi=0.42$ | $\phi=0.5$ | $\phi=0.58$ | $\phi=0.42$ | $\phi=0.5$ | $\phi=0.58$ |
| $\mathcal{M}_0$ | 14 | 14 | 14 | 14 | 14 | 14 |
| $\mathcal{M}_1$ | 5 | 5 | 5 | 6 | 4 | 5 |
| $\mathcal{M}_2$ | 4 | 4 | 4 | 7 | 5 | **3** |
| $\mathcal{M}_3$ | 11 | 11 | 12 | **1** | 11 | 13 |
| $\mathcal{M}_4$ | **1** | **2** | **2** | **2** | **1** | 6 |
| $\mathcal{M}_5$ | **2** | **1** | **1** | 4 | **2** | **2** |
| $\mathcal{M}_6$ | **3** | **3** | **3** | 5 | **3** | 4 |
| $\mathcal{M}_7$ | 13 | 13 | 13 | 13 | 13 | 12 |
| $\mathcal{M}_8$ | 8 | 8 | 6 | 10 | 8 | 7 |
| $\mathcal{M}_9$ | 7 | 7 | 8 | 9 | 6 | 8 |
| $\mathcal{M}_{10}$ | 6 | 6 | 7 | 8 | 7 | 9 |
| $\mathcal{M}_{11}$ | 9 | 9 | 9 | 11 | 9 | **1** |
| $\mathcal{M}_{12}$ | 10 | 10 | 10 | **3** | 10 | 10 |
| $\mathcal{M}_{13}$ | 12 | 12 | 11 | 12 | 12 | 11 |

*Notes*: entries represent the ranking of models based on the statistics $\sqrt{P^{-1}\sum_{t=1}^{P}\mathcal{L}_t}$, where $\mathcal{L}_{\sqcup}$ is the flexible loss function in Eq. (1) and $P=351$; statistics in panel (a) - (c) are based on the univariate loss for forecast horizons $h=1,7,28$, while those in panel (d) are based on the multivariate loss for $h=1,...,28$; the shape of the loss function is determined by $\rho=2$ and the asymmetry coefficient: $\phi=0.42,0.50,0.58$ ($\tau=-0.16,0.00,0.16$); these values guarantee that the multivariate loss is always non-negative (see Komunjer and Owyang (2012) for details). Models, $\mathcal{M}_m$, are described in Table 1. Entries in bold denote models with the lowest loss.

## 4.2 Additional call arrival data

The relevance of our results clearly depends on whether they can be extended to other call arrival data. To shed light on this issue, we consider two additional series representing the number of call arrivals recorded at the call centers operated by two banks, one located in Israel and the other in the U.S.. In Table 9 we list, for each series, the model with the lowest MSFE. Our results show that the ARMAX–GARCH and the SARMAX–GARCH models are the winning options in all countries and at all forecast horizons.

Table 9: MSFE ranking of forecasting models for call centers in Israel, Italy and the U.S.

| country | $h = 1$ | $h = 7$ | $h = 28$ | $h = 1, ..., 28$ |
|---|---|---|---|---|
| Israel | $\mathcal{M}_5$ | $\mathcal{M}_5$ | $\mathcal{M}_2$ | $\mathcal{M}_5$ |
| Italy | $\mathcal{M}_7$ | $\mathcal{M}_5$ | $\mathcal{M}_5$ | $\mathcal{M}_5$ |
| U.S. | $\mathcal{M}_2$ | $\mathcal{M}_2$ | $\mathcal{M}_3$ | $\mathcal{M}_2$ |

*Notes*: the first column identifies the dataset used; entries in columns 2-4 represent the model with the lowest Mean Squared Forecast Error (MSFE) at forecast horizon $h = 1$, 7, 28 days, while entries in the last column represent the model with the lowest sum of MSFE over forecast horizons $h = 1, 2, \ldots, 28$. Models are described in Tables 1.

## 4.3 Degree of asymmetry of the loss function

In Section 3 the degree of asymmetry has been restricted to lie in the range $\phi = (0.42, 0.58)$ because, with $h = H = 28$, the multivariate loss is non–negative for $|\tau| < 1/\sqrt{H}$, that is $|\tau| < 0.18$. This corresponds to $\phi = (|\tau| + 1)/2$, that is 0.41 and 0.59. Therefore, while in the multivariate case we cannot vary $\phi$ any further, when evaluating one forecast horizon at the time we can consider more extreme values of $\phi$. Figure 2 shows the best–performing model for $\phi = 0.1, 0.2, \ldots, 0.9$ and $h = 1, 7, 28$. The ranking remains surprisingly stable for a wide range of values of the asymmetry parameter.

## 4.4 On the usefulness of second–moment modeling

We have highlighted that the addition of a GARCH component to ARMAX and SARMAX models improves their forecasting performance. However, having focused on point forecasts, we have not fully exploited the strength of these specifications, namely the modeling of the conditional variance of call arrival data. The GARCH component can be exploited in the construction of optimal point forecasts for log–transformed variables and for density forecasts.

Figure 2: Ranking of models as a function of the asymmetry of the loss function

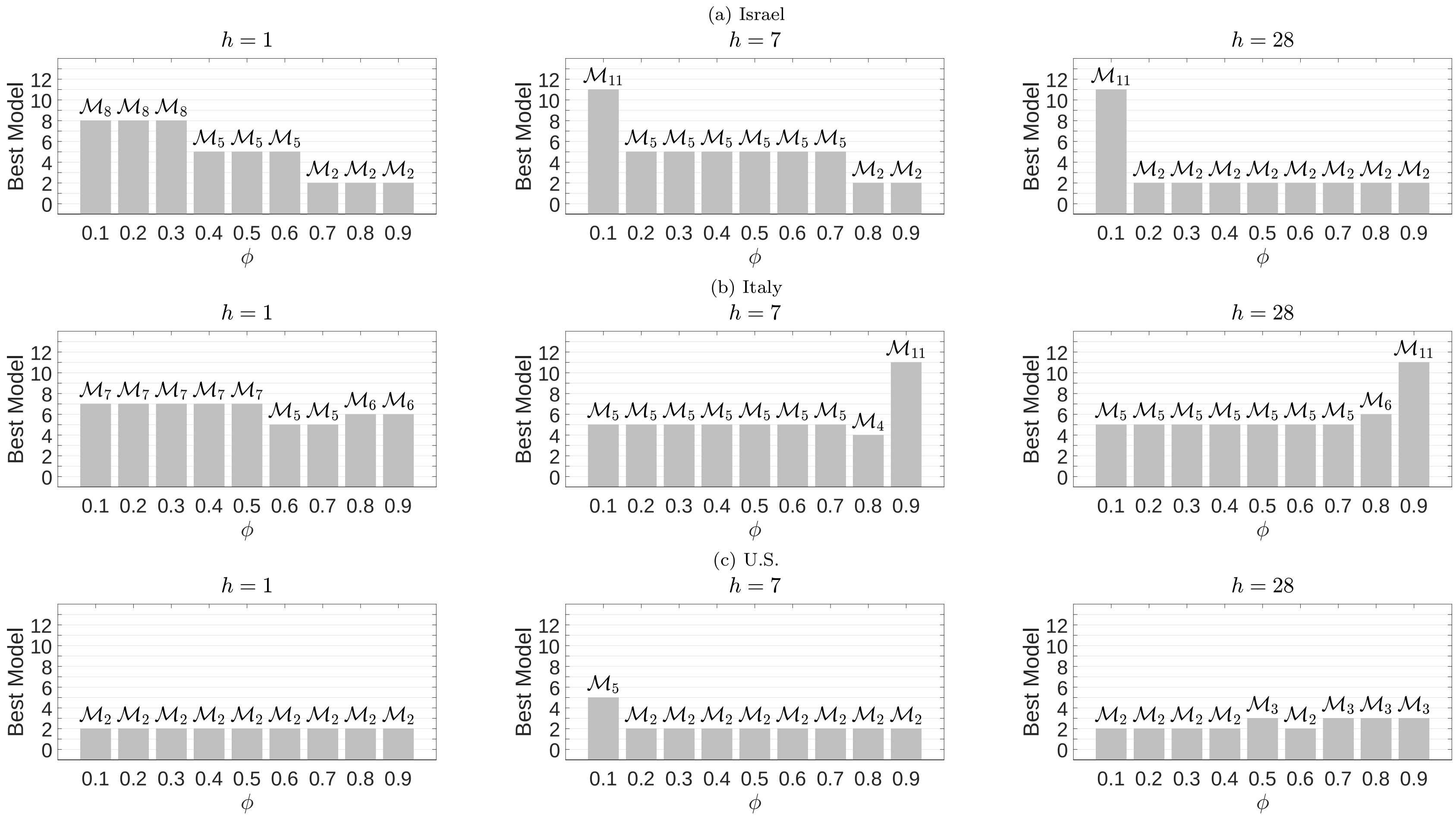

*Notes:* the figure shows the model that minimizes the loss function for different values of the asymmetry degree $\phi = 0.1, 0.2, 0.3, 0.4, 0.5, 0.6, 0.7, 0.8, 0.9$ ($\tau = 2 \times \phi - 1 =$ -0.8, -0.6, -0.4, -0.2, 0.0, 0.2, 0.4, 0.6, 0.8)

Table 10: Naive vs optimal forecasts

| Model | Israel | Italy | US |
|---|---|---|---|
| $\mathcal{M}_1$ | **0.7477** | 1.0455 | **0.9802** |
| $\mathcal{M}_2$ | **0.9778** | 1.0378 | **0.9867** |
| $\mathcal{M}_4$ | **0.7133** | 1.0380 | **0.9802** |
| $\mathcal{M}_5$ | **0.9758** | 1.0300 | **0.9861** |

*Notes:* entries are the ratio of the Root Mean Squared Forecast Error of optimal and naive one-step ahead forecasts: $RMSFE^{opt}/RMSFE^{naive}$. Cases when the optimal forecast outperforms the naive forecast are indicated with bold text.

*Naive vs optimal point forecasts.* Since our results indicate that forecasts tend to improve when a GARCH component is added to the model, it is worth investigating the usefulness of second-moment modeling in detail. ARMAX and SARMAX models have been estimated after log-transforming call arrival data ($Y_t$) and forecasts are obtained by the inverse of the log transformation: $f_{t+h|t} = \exp\left(y_{t+h|t}\right)$, where $y_{t+h|t} = E\left(y_{t+h}\left|y_t, y_{t-1}, ...\right.\right)$ and $y_t = log(Y_t)$. Granger and Newbold (1976) label this approach as "naive". The naive forecast is biased, whereas an "optimal" forecast that uses the forecast error variance of the transformed series ($\sigma_y^2$) to correct the bias would be $f^{opt}_{t+h|t} = \exp\left(y_{t+h|t} + 0.5 \times \sigma_y^2(h)\right)$.

Therefore, we compare ARMA and SARMAX models, with and without the GARCH component, to see if incorporating the conditional variance predictions in the optimal forecast formula yields to more accurate point forecasts than simply relying on the naive approach. In the case of ARMAX and SARMAX models without GARCH component, we have used the variance of the residuals as an estimate of $\sigma_y^2$. For the sake of brevity, we have carried out this exercise only for one-step ahead forecasts, evaluated with the symmetric MSE loss. Table 10 shows that optimal forecasts are better than naive forecasts in eight comparisons out of twelve. Moreover, improvements associated with models with a GARCH component are generally smaller than models without such component. These results are in line with those of Lütkepohl and Xu (2010), who show that the accuracy gains due to use of optimal instead of naive forecasts are usually minimal.

*Density forecasts.* Density forecasts have been produced for a subset of models that include the benchmark SRW, ARMAX and SARMAX with and without GARCH component, the Poisson and the Negative Binomial specifications.

Table 11: Density and quantile forecasts evaluation

| Israel | | | | | |
|---|---|---|---|---|---|
| Model | RPS | 5% | 25% | 75% | 95% |
| $\mathcal{M}_0$ | **415.75** | 22.83 | 81.52 | 100.00 | 100.00 |
| $\mathcal{M}_1$ | 660.51 | 10.87 | 43.48 | 100.00 | 100.00 |
| $\mathcal{M}_2$ | 763.86 | 6.52 | 36.96 | **88.04** | 97.83 |
| $\mathcal{M}_4$ | 707.10 | 9.78 | 44.57 | 98.91 | 100.00 |
| $\mathcal{M}_5$ | 752.42 | 7.61 | **32.61** | 89.13 | **95.65** |
| $\mathcal{M}_8$ | 1650.92 | 2.17 | 5.43 | 13.04 | 26.09 |
| $\mathcal{M}_9$ | 835.89 | **4.35** | 38.04 | 94.57 | 100.00 |
| Italy | | | | | |
| Model | RPS | 5% | 25% | 75% | 95% |
| $\mathcal{M}_0$ | 4604.45 | 6.15 | 36.00 | 87.69 | 98.46 |
| $\mathcal{M}_1$ | 3966.79 | 7.08 | 41.85 | 89.54 | 97.23 |
| $\mathcal{M}_2$ | 4019.20 | **5.85** | 36.00 | **82.46** | **94.46** |
| $\mathcal{M}_4$ | **3941.04** | 8.31 | 42.77 | 89.23 | 96.92 |
| $\mathcal{M}_5$ | 4009.09 | 7.08 | **34.15** | 82.77 | **94.46** |
| $\mathcal{M}_8$ | 11174.43 | 0.00 | 0.62 | 2.77 | 6.15 |
| $\mathcal{M}_9$ | 4800.73 | 7.69 | 40.92 | 84.31 | 96.00 |
| U.S. | | | | | |
| Model | RPS | 5% | 25% | 75% | 95% |
| $\mathcal{M}_0$ | 6836.74 | 26.94 | 78.89 | 86.39 | 89.17 |
| $\mathcal{M}_1$ | 4132.72 | 6.11 | 35.28 | 81.39 | 92.50 |
| $\mathcal{M}_2$ | 3487.27 | **5.83** | 27.22 | **75.83** | 90.00 |
| $\mathcal{M}_4$ | 4106.38 | 6.11 | 36.39 | 82.22 | 92.50 |
| $\mathcal{M}_5$ | **3485.90** | 6.11 | 28.06 | 76.39 | 90.83 |
| $\mathcal{M}_8$ | 10224.19 | 0.00 | 0.56 | 2.22 | 3.61 |
| $\mathcal{M}_9$ | 5048.43 | 3.89 | **24.44** | 77.22 | **95.56** |

*Notes:* column 2 presents Ranked Probability Score (RPS), while columns 3-6 show the Empirical Coverage Probability (ECP) of 5%, 25%, 75% and 95% prediction intervals. In all cases the most accurate model is identified with boldface font.

We compare density forecasts obtained from count data models with those from linear specifications on log–transformed data, assuming that the series in levels are log-normal. One–step–ahead density forecasts are computed with Monte Carlo simulations. For each observation in the evaluation sample we have drawn 1000 observations from the relevant distribution using the estimated or forecast parameters. In particular, for SRW, ARMA and SARMAX models, we simulate from a log–normal distribution with parameters equal to the one–step ahead forecast of its mean and variance. When the model has a GARCH component, we rely of the forecast of the conditional variance, otherwise we use the estimate of the variance of the residuals. In the case of the Poisson, we rely on the one–step–ahead estimate of the arrival rate. Lastly, for the Negative Binomial model, density forecasts are drawn using the in-sample estimate of the over–dispersion parameter.

Density forecasts are evaluated using the Ranked Probability Score (RPS):

$$RPS_t = \sum_{j=0}^{J} [F_t(j) - I(Y_t \leq j)]^2 \tag{7}$$

where $F_t$ is the forecast of the empirical cumulative distribution function of $Y_t$, $I(.)$ is the indicator function and $J$ is the possible maximum number of calls at date $t$, which is set as the maximum between the realized and largest simulated forecast of $Y_t$. Averaging over the evaluation sample we get: $\overline{RPS} = H^{-1} \sum_{t=1}^{H} RPS_t$, which is a loss function and it is lower for models that perform better in terms of density forecast ability.

Moreover, we evaluated distributional forecasts computing the Empirical Coverage Probability (ECP) of 5%, 25%, 75% and 95% prediction intervals:

$$ECP_\theta = \frac{1}{H} \sum_{t=1}^{H} I\left(q_t^{0.5\theta} < Y_t < q_t^{1-0.5\theta}\right) \tag{8}$$

Results in Table 11 show that ARMAX and SARMAX models yield the lowest RPS for both the Italian and U.S. call arrival data, while in the case of Israeli, the SRW is the preferred option. When looking at the ECP of the prediction interval we can see that in 9 cases out of 12 the most accurate forecasts are produced with models including a GARCH component. Interestingly, there is also evidence that modeling the over–dispersion of call arrival data with a Negative Binomial model might be a useful approach for density forecasting.

# 5 Conclusions

Call centers' managers and companies relying on call center services are interested in obtaining accurate forecasts of call arrivals for achieving optimal operating efficiency. This paper has shown how to choose among competing forecasting methods for call arrivals in call centers.

The empirical exercise in this paper mimics the interaction between a professional forecaster and a manager who uses forecasts of incoming calls to decide how many operators are required each day at a call center. In this context, we have evaluated fourteen models and a set of seven forecast combination schemes using flexible loss functions, statistical tests and economic measures of performance.

Each of the forecast models discussed in this paper is able to capture one or more key features of the daily call arrival series. Moreover, all of the models and combination methods are computationally tractable, with a relatively small number of parameters that can be easily estimated and updated with any off–the–shelf statistical software as new data become available. We view simplicity as an important point in the selection of a model for forecasting call arrivals. In fact, to be of practical use, a forecast model not only should reproduce the key features of the data, but also it has to be easily implementable, and able to quickly generate new forecasts to support the operational decisions in call centers (Ibrahim et al., 2016; Mehrotra and Fama, 2003).

We have shown that the professional forecaster can reduce the number of proposed models using the Reality Check test and the Model Confidence Set. These tests, as well as models ranking, suggest that the best available options are the SARMAX–GARCH and a combined forecast obtained with ABMA. Subsequently, we have shown that the economic evaluation of forecast accuracy leads to the same selection. However, given that individual and combined forecasts have approximately the same monetary value, the manager will choose eventually the SARMAX–GARCH model, due to lower maintenance costs. The maintenance costs of a forecasting model used by an employee of the call center include direct costs, related to periodical checks of the specification, as well as indirect costs, associated to the relative complexity of the forecasting method. Given that forecast combinations require a set of models, whose specifications have to be periodically checked, this approach will probably lead to higher maintenance costs with respect to the SARMAX–GARCH model, which is then the best available choice. Moreover, combined forecasts cannot be used for simulation purposes, which typically need a unique set of key parameters.

Since our paper relies on a wide array of results covering different loss functions, forecast horizons and call arrival series, a number of more comprehensive conclusions can be drawn. First, since the benchmark SRW model is always outperformed by other relatively more general specifications, investing in the use of these models might be a good option for the management of the call center. Second, irrespective of the forecast horizon and the shape of the loss function, forecast combination, especially if based on optimal combining weights calculated by means of ABMA, proves to be useful and leads to lower statistical losses than most forecasts obtained from individual models. Third, the statistical evaluation of each forecast model indicates that second-moment modeling, in addition to seasonality, is important. In fact, the ARMAX–GARCH and SARMAX–GARCH models emerge as the best alternatives among both individual and combined forecasts. This last result implies that anticipating the variability of call arrivals is extremely relevant since, when a call center operates under a SLA, higher uncertainty requires higher staffing levels to meet service quality objectives. Moreover, in the light of its ease of implementation, the SARMAX–GARCH model seems to be a good candidate also for density forecasting.

# Appendix to

# "Statistical and Economic Evaluation of Time Series Models for Forecasting Arrivals at Call Centers"

# Contents

# A Flexible loss functions

In this section we present more details about the loss function used in the paper. The statistical evaluation of forecasts is based on the loss function put forth by Komunjer and Owyang (2012), that is a multivariate generalization of the loss function due to Elliott et al. (2005, 2008). The strength of these metrics is to encompass a variety of symmetric and asymmetric loss functions that are often used in empirical applications.

As for the notation, let $f_{i,t+h|t} \equiv E(Y_{t+h}|\mathcal{I}_t)$ be the $h$-step ahead forecast of $Y_{t+h}$ issued with model $i$ ($\mathcal{M}_i$), where $\mathcal{I}_t$ is the information set available at time $t$ and $h$ is the forecast horizon. The corresponding forecast error is $u_{i,t+h|t} = Y_{t+h} - f_{i,t+h|t}$. The model subscript $i$ is dropped for ease of notation, and a set of $(P \times 1)$ vectors of forecasts errors, one for each forecast horizon, is considered: $\mathbf{u}_{(h)} = [u_{(h)1}, \ldots, u_{(h)P}]'$ for $h = 1, \ldots, H$, where $H$ is the maximum forecast horizon (i.e. 28 days in this paper) and $P$ is the size of the evaluation sample. These vectors form the rows of a $(H \times P)$ matrix: $\mathbf{u} = [\mathbf{u}_{(1)} \ldots \mathbf{u}_{(H)}]'$, where $\mathbf{u}_p$, $p = 1, \ldots, P$, indicates its $p$-th column. With this notation, we can write the multivariate loss function of Komunjer and Owyang (2012) as:

$$\mathcal{L}_p\left(\mathbf{u}_p; \rho, \boldsymbol{\tau}\right) = \left(\left\|\mathbf{u}_p\right\|_\rho + \boldsymbol{\tau}'\mathbf{u}\right) \left\|\mathbf{u}_p\right\|_\rho^{\rho-1} \tag{1}$$

where $\left\|\mathbf{u}_p\right\|_\rho \equiv \left(\sum_{h=1}^{H} \left|u_{(h)p}\right|^\rho\right)^{\frac{1}{\rho}}$.

*Properties of the multivariate loss function.* The properties of the multivariate loss function in equation (1) are established by Komunjer and Owyang (2012). The elements of the $(H \times 1)$ vector $\boldsymbol{\tau}$ determine the degree of asymmetry of the loss function. We assume that all the elements of $\boldsymbol{\tau}$ are equal to $\tau$, so that the degree of asymmetry is the same at all forecast horizons $h$, although we might consider a different $\tau$ for each $h$. Moreover, $-1 \leq \tau \leq 1$, so that the loss is symmetric and additively separable for $\tau = 0$. Lastly, it is useful to recall that the non-negativeness of $\mathcal{L}_p$ is ensured if the $q$ norm of $\boldsymbol{\tau}$ is less than unity, where $\frac{1}{p} + \frac{1}{q} = 1$, $1 \leq p \leq \infty$, with the convention that $q = \infty$ if $p = 1$ and $\left\|\boldsymbol{\tau}\right\|_\infty = \max\{\tau_1, \ldots, \tau_H\}$. For $\rho = 2$ and with a common $\tau$ for all $h = 1, \ldots, H$, the non-negativeness of $\mathcal{L}_p$ requires $\left\|\boldsymbol{\tau}\right\|_2 < 1$, which simplifies to $|\tau|\sqrt{H} < 1$ and hence $|\tau| < \frac{1}{\sqrt{H}}$.

For example, when $H = 1, 2, 28$, we need $|\tau| < 1$, $|\tau| < 0.7$ and $|\tau| < 0.18$, respectively.

## A.1 Univariate loss

To fix the ideas, let us consider the case when the forecaster wants to evaluate forecasts at a single horizon, so that $H = 1$ and $\mathbf{u} = \mathbf{u}'_{(1)} = [u_{(1)1}, \ldots, u_{(1)P}]$. In this case we can simplify the notation further and write $\mathbf{u} = [u_1, \ldots, u_P]$. Thus, since the $p$-th column of $\mathbf{u}$ is a scalar, we have: $\|u_p\|_\rho \equiv (|u_p|^\rho)^{\frac{1}{\rho}} = |u_p|$. Therefore, equation (1) reduces to:

$$\begin{aligned}
\mathcal{L}_p(u_p; \rho, \tau) &= (|u_p| + \tau \times u)\, |u_p|^{\rho-1} \\
&= \left[1 + \tau \times \frac{u}{|u_p|}\right] \times |u_p|^\rho \\
&= [1 + \tau \times sign(u_p)] \times |u_p|^\rho \\
&= [1 + \tau \times (2I(u_p > 0) - 1)] \times |u_p|^\rho \qquad (2)
\end{aligned}$$

where $I(.)$ is the indicator function: $I(u_p > 0) = 1$ if $u_p > 0$, $I(u_p > 0) = 0$ if $u_p < 0$, $I(u_p > 0) = \frac{1}{2}$ if $u_p = 0$.[1] Then, letting $\tau = 2\phi - 1$, we can write equation (2) as:

$$\begin{aligned}
\mathcal{L}_p(u_p; \rho, \tau) &= [1 + \tau \times (2I(u_p > 0) - 1)] \times |u_p|^\rho \\
&= [1 + (2\phi - 1) \times (2I(u_p > 0) - 1)] \times |u_p|^\rho \\
&= 2\,[1 - \phi + (2\phi - 1) \times I(u_p > 0)] \times |u_p|^\rho \\
&= 2\,[1 - \phi + \tau \times I(u_p > 0)] \times |u_p|^\rho \qquad (3)
\end{aligned}$$

Finally, since $I(u_p > 0) = 1 - I(u_p < 0)$, we have:

$$\begin{aligned}
\mathcal{L}_p(u_p; \rho, \tau) &= 2\,[1 - \phi + \tau \times I(u_p > 0)] \times |u_p|^\rho \\
&= 2\,[1 - \phi + (2\phi - 1) \times (1 - I(u_p < 0))] \times |u_p|^\rho \\
&= 2\,[\phi + (1 - 2\phi) \times I(u_p < 0)] \times |u_p|^\rho \qquad (4)
\end{aligned}$$

Notice that expression (4) corresponds to the univariate flexible loss of Elliott et al. (2005).

[1] To derive equation (2) we have used the facts that $sign(x) = \frac{x}{|x|}$ for $x \neq 0$ and that $sign(x) = 2I(u > 0) - 1$.

The loss function is asymmetric for $\phi \neq 0.5$. In particular, over-forecasting (negative forecast errors) is costlier than under-forecasting for $\phi < 0.5$. On the contrary, when $\phi > 0.5$, positive forecast errors (under-prediction) are more heavily weighed than negative forecast errors (over-prediction).

*Special cases.* Figure A1 shows that equation (4) nests several special cases. For $\rho = 1$ equation (4) gives the lin-lin loss:

$$\mathcal{L}_p\left(u_p; 1, \phi\right) = 2\left[\phi + (1 - 2\phi) \times I(u_p < 0)\right] \times |u_p| \tag{5}$$

Optimal forecasts from the lin-lin loss are conditional quantiles (see Koenker and Bassett, 1978; Gneiting, 2011). The function is symmetric and boils down to the Mean Absolute Error loss for $\phi = 0.5$ ($\tau = 0$):

$$\mathcal{L}_p\left(u_p; 1, \frac{1}{2}\right) = |u_p| \tag{6}$$

For $\rho = 2$ equation (4) gives the quad-quad loss:

$$\mathcal{L}_p\left(u_p; 2, \phi\right) = 2\left[\phi + (1 - 2\phi) \times I(u_p < 0)\right] \times |u_p|^2 \tag{7}$$

Optimal forecasts from the quad-quad loss are expectiles (Newey and Powell, 1987). See Bastianin et al. (2014) for the use of expectiles in a forecast horse-race. The function is symmetric and boils down to the Mean Squared Error loss for $\phi = 0.5$ ($\tau = 0$):

$$\mathcal{L}_p\left(u_p; 2, \frac{1}{2}\right) = u_p^2 \tag{8}$$

Figure A1: Examples of flexible loss functions

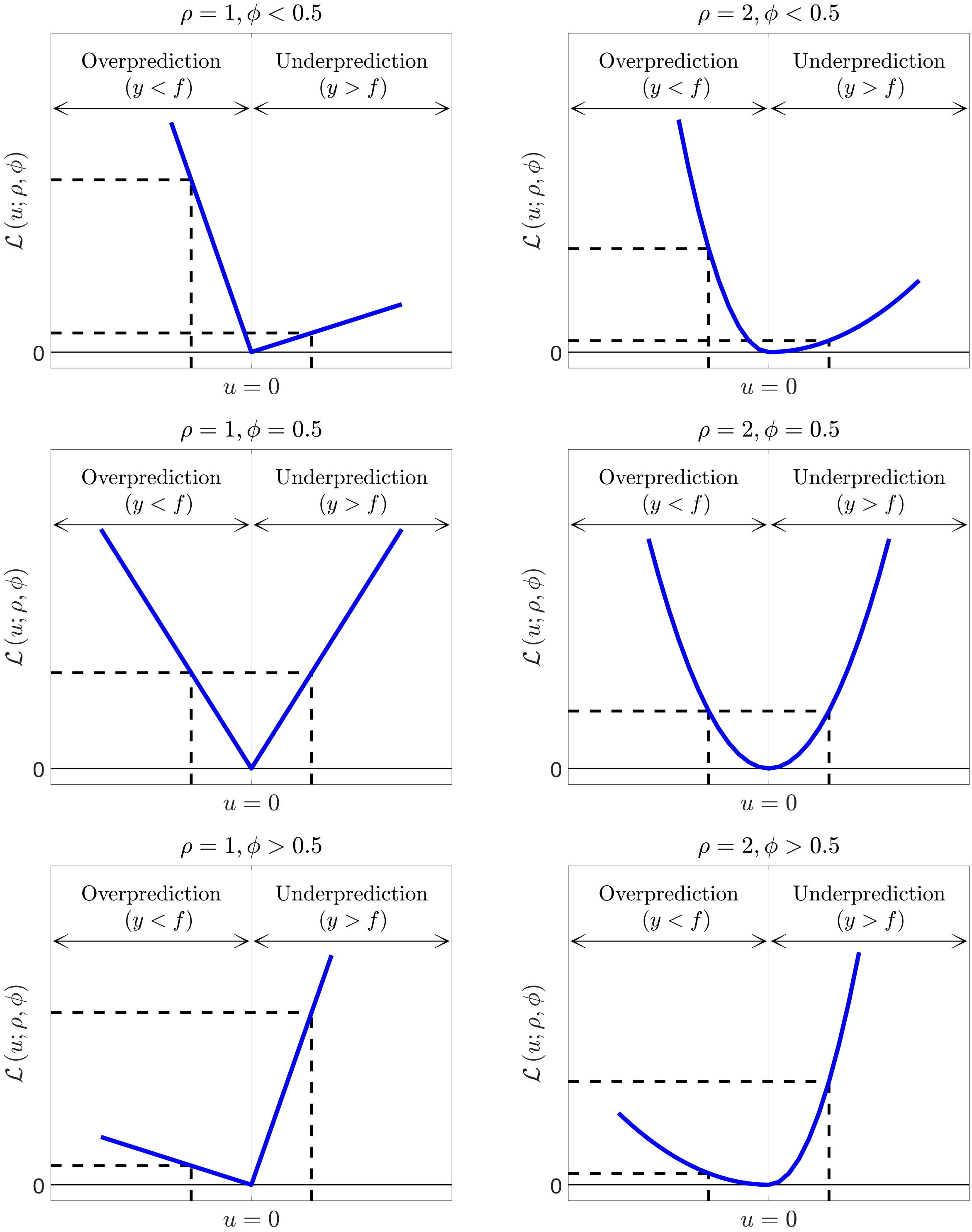

*Notes:* The figure shows the flexible loss function of Elliott et al. (2005) for different $\rho$ and $\phi$. The flexible loss function is given by: $\mathcal{L}_p(u_p; \rho, \phi) = 2\left[\phi + (1 - 2\phi) \times I(u_p < 0)\right] \times |u_p|^\rho$, where $u = y - f$ are forecast errors and $\rho \geq 1$ and $0 < \phi < 1$. The first column of plots shows lin-lin losses ($\rho = 1$) for different asymmetry levels ($\phi$). The second column of plots shows quad-quad losses ($\rho = 2$) for different asymmetry levels ($\phi$). Plots in the first row show that over-prediction (i.e. $y - f < 0$) is costlier than under-prediction (i.e. $y - f > 0$) for $\phi < 0.5$. Plots in the second row show that the loss is symmetric about forecast errors when $\phi = 0.5$. Plots in the third row show that under-prediction is costlier than over-prediction for $\phi > 0.5$.

## A.2 Bivariate loss

Let us now move to the case $H = 2$, so that there are only two series of forecast errors, one- and two-step ahead. In this case $\mathbf{u} = [\mathbf{u}_{(1)},\ \mathbf{u}_{(2)}]'$ is a $(2 \times P)$ matrix with the $p$-th column given by $\mathbf{u}_p = [u_{(1)p},\ u_{(2)p}]'$ for $p = 1, \ldots, P$. Assuming that $\rho = 2$, the multivariate loss function in equation (1) reduces to:

$$\begin{aligned}
\mathcal{L}_p\left(\mathbf{u}_p; 2, \boldsymbol{\tau}\right) &= \left(\left\|\mathbf{u}_p\right\|_2 + \boldsymbol{\tau}'\mathbf{u}\right)\left\|\mathbf{u}_p\right\|_2 \\
&= \left(\left(\sum_{h=1}^{2} u_{(h)p}^2\right)^{\frac{1}{2}} + \tau u_{(1)p} + \tau u_{(2)p}\right)\left(\sum_{h=1}^{2} u_{(h)p}^2\right)^{\frac{1}{2}} \\
&= \left(\left(u_{(1)p}^2 + u_{(2)p}^2\right)^{\frac{1}{2}} + \tau u_{(1)p} + \tau u_{(2)p}\right)\left(u_{(1)p}^2 + u_{(2)p}^2\right)^{\frac{1}{2}} \\
&= \left(u_{(1)p}^2 + u_{(2)p}^2\right) + \tau \times \left(u_{(1)p} + u_{(2)p}\right) \times \sqrt{\left(u_{(1)p}^2 + u_{(2)p}^2\right)} \qquad (9)
\end{aligned}$$

As in the univariate case, the bivariate (and, more generally, the multivariate) loss includes some special cases: when $\tau = 0$ ($\phi = 0.5$) and $\rho = 2$, we obtain the trace of the MSE loss function, while for $\tau = 0$ ($\phi = 0.5$) and $\rho = 1$, equation (9) reduces to the trace of the MAE loss function (see Zeng and Swanson, 1998). In both cases, symmetry also ensures the multivariate loss to be additively separable in univariate losses. On the contrary, when $\tau \neq 0$ ($\phi \neq 0.5$), the loss function becomes asymmetric and is not additively separable in individual losses. This is illustrated for the bivariate case in Figure A2.

Figure A2: Bivariate flexible loss function and sum of individual losses: contour plots

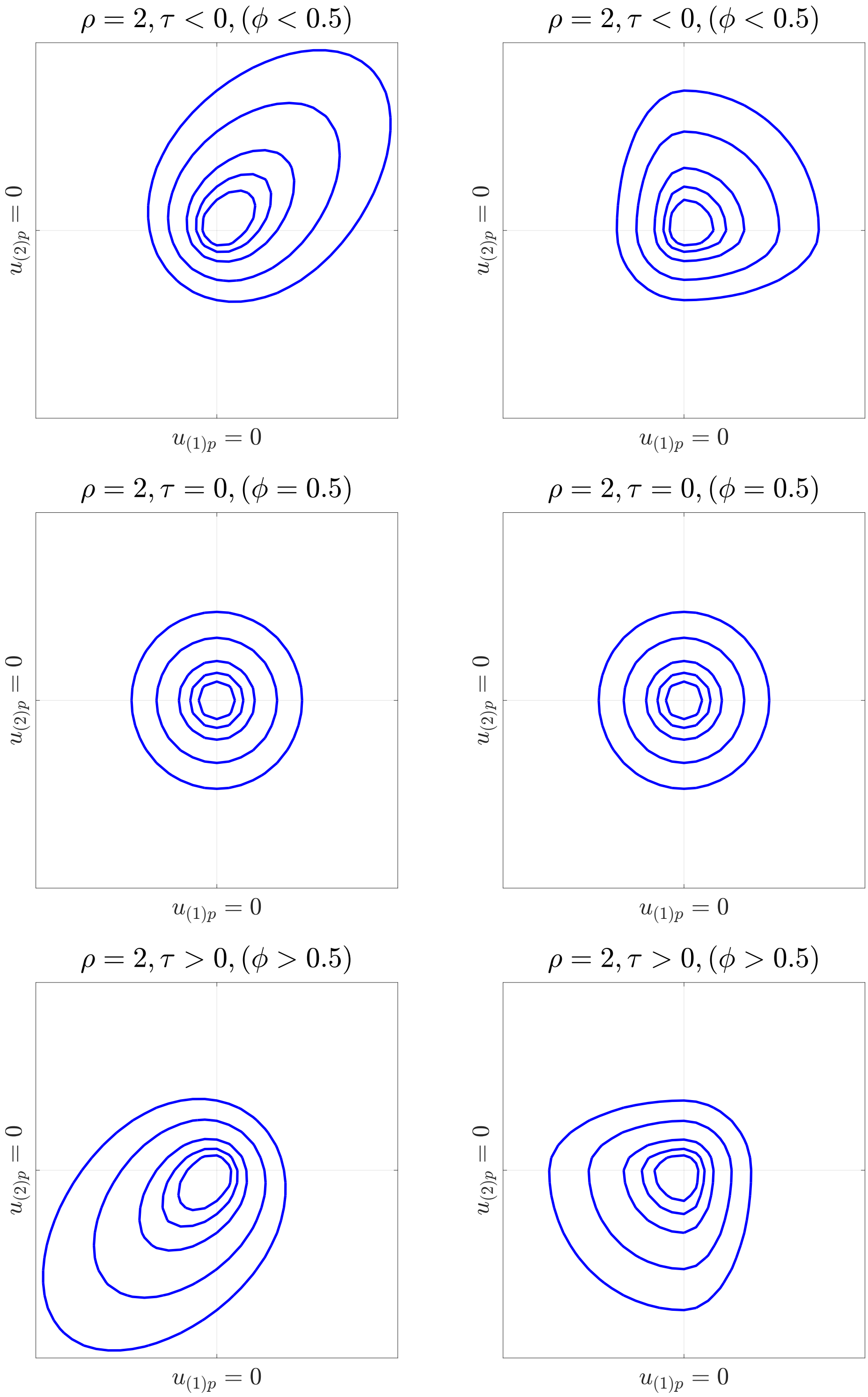

*Notes:* figures in the left panel show the contour plots of the bivariate flexible loss functions of Elliott et al. (2005) for different $\rho$ and $\tau$. The right panel shows the iso-loss curves obtained as the sum of univariate flexible losses with the same $\rho$ and $\tau$. $u = y - f$ are forecast errors and $\tau = 2\phi - 1$. Plots in the first row show that over-prediction (i.e. $y - f < 0$) is costlier than under-prediction (i.e. $y - f > 0$) for $\tau < 0$ ($\phi < 0.5$). Plots in the second row show that the loss is symmetric about forecast errors when $\tau = 0$ ($\phi = 0.5$). Plots in the third row show that under-prediction is costlier than over-prediction for $\tau > 0$ ($\phi > 0.5$).

# B Descriptive statistics

## B.1 Additional figures and descriptive statistics

Table B1 shows the sample average of the three series of daily call arrivals used in the paper. It is evident from the second column that the Italian and the U.S. series are characterized by very high daily call volumes, while the Israeli series consists of lower counts. The third and fourth columns report the coefficient of variation of the series and of their log-transformations. In all cases the logarithm seems to stabilize the variance of the series.

Figures B3, B4, B5 show the line plots of daily call arrivals series used in the paper and the boxplots of their distribution over the days of the week and the months of the year.

Figure B6 shows the log-transformed daily call arrivals and fitted values from cubic spline interpolation. The estimated autocorrelation of the three series are shown in Figure B7.

Table B1: Descriptive statistics for daily call arrivals: Israel, Italy and the U.S.

| Country | Avg. $Y_t$ | C.V. $Y_t$ | C.V. $\log Y_t$ | $T$ |
|---|---|---|---|---|
| Israel | 960 | 0.51930 | 0.14027 | 361 |
| Italy | 31258 | 0.60040 | 0.08743 | 749 |
| U.S. | 45145 | 0.43377 | 0.06117 | 893 |

*Notes*: the first column shows the sample average daily call arrivals ($Y_t$), the second column reports the coefficient of variation of $Y_t$ (C.V. $Y_t$), the third column shows the coefficient of variation of $\log Y_t$ (C.V. $\log Y_t$). The coefficient of variation, as well as the sample average, are calculated using the a whole sample of observations, whose size is shown in the fourth column.

Figure B3: Daily call arrivals for the Italian call center

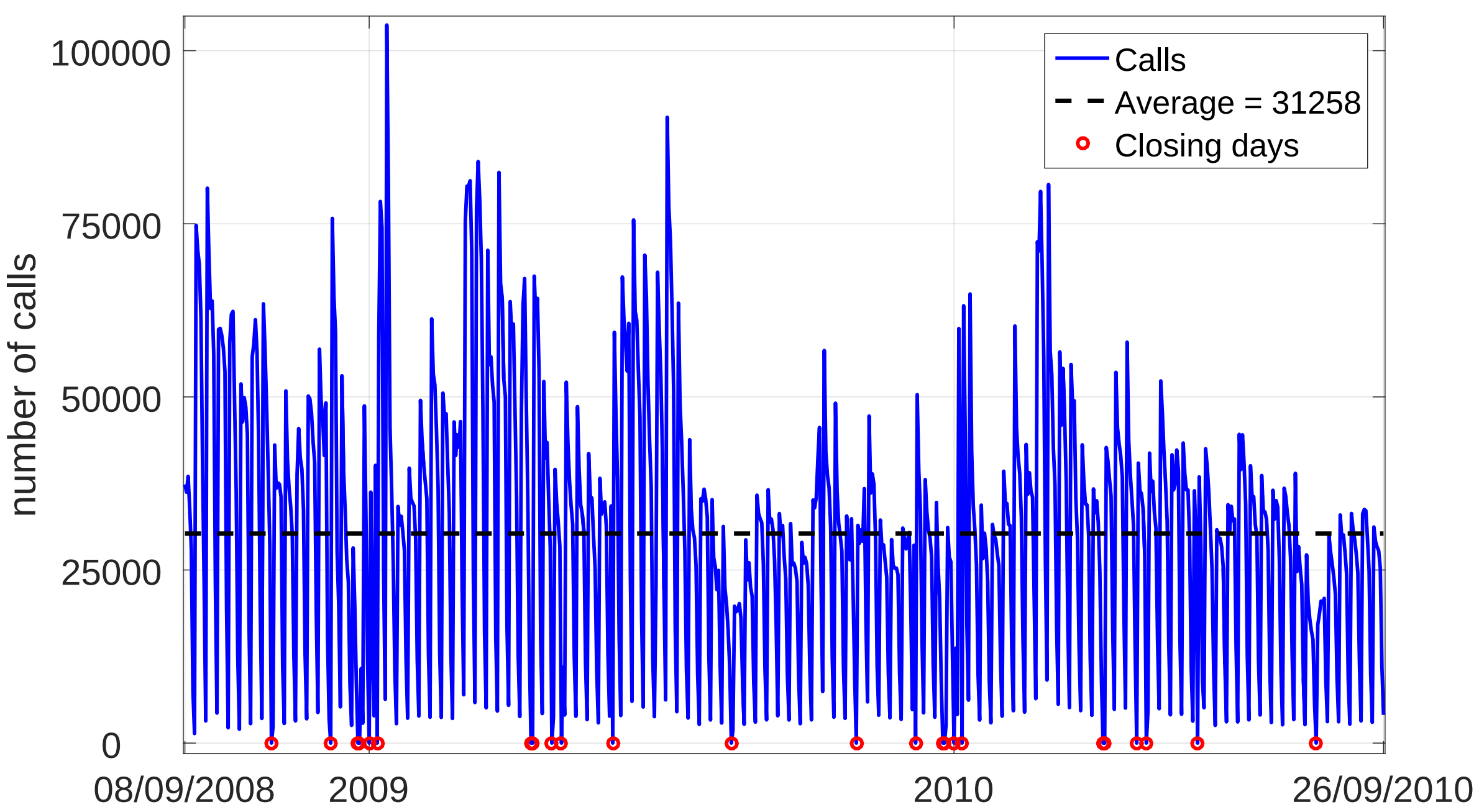

(a) Daily call arrivals

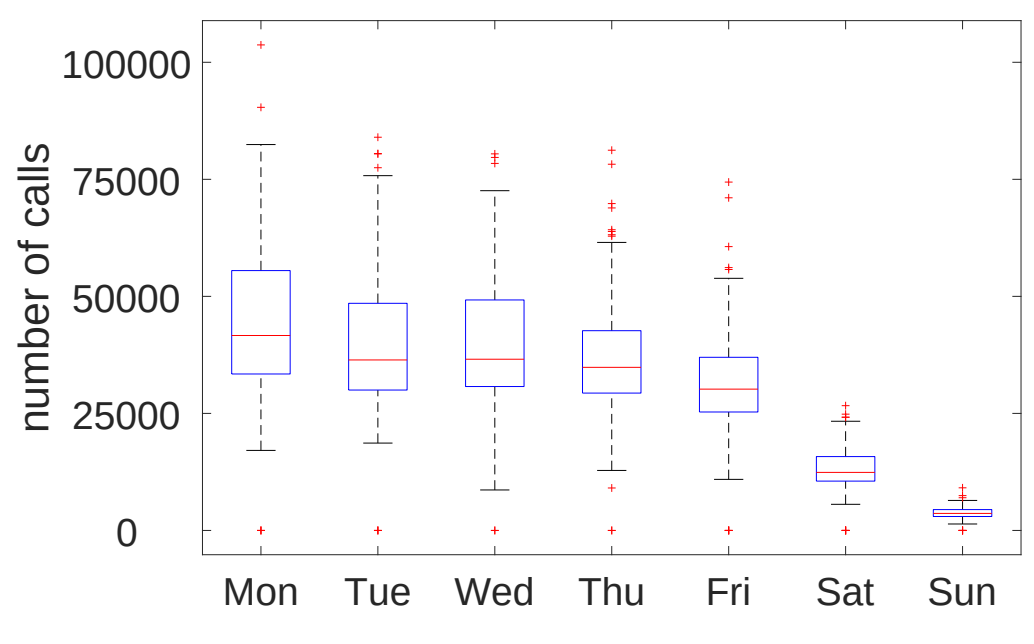

(b) Box plots of daily call arrivals per day for each day of the week.

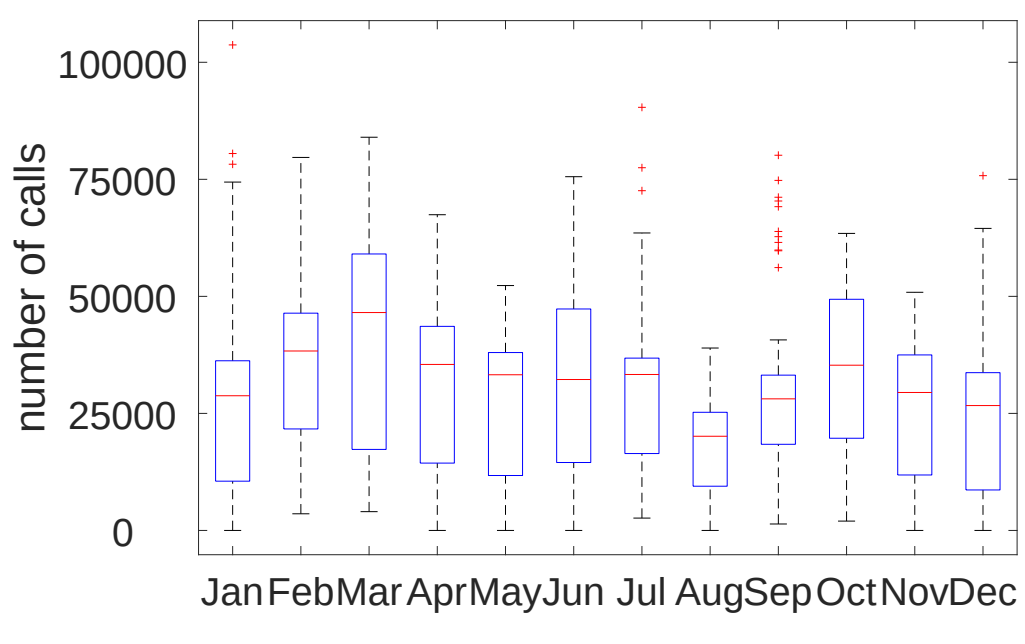

(c) Box plots of daily call arrivals per month for each month of the year.

*Notes:* panel (a) shows the daily call arrival series (continuous line), its sample average (horizontal dashed line) and closing days (circles). In each box plot in panels (b) and (c), the central mark indicates the median, the bottom and top edges of the box indicate the 25th and 75th percentiles, respectively. The two bars, known as whiskers, are located at a distance of 1.5 times the interquartile range (i.e. the height of the box) below the 25th percentile and above the 75th percentile. Daily observations that fall above or below these two bars are denoted with crosses.

Figure B4: Daily call arrivals for the Israeli call centre

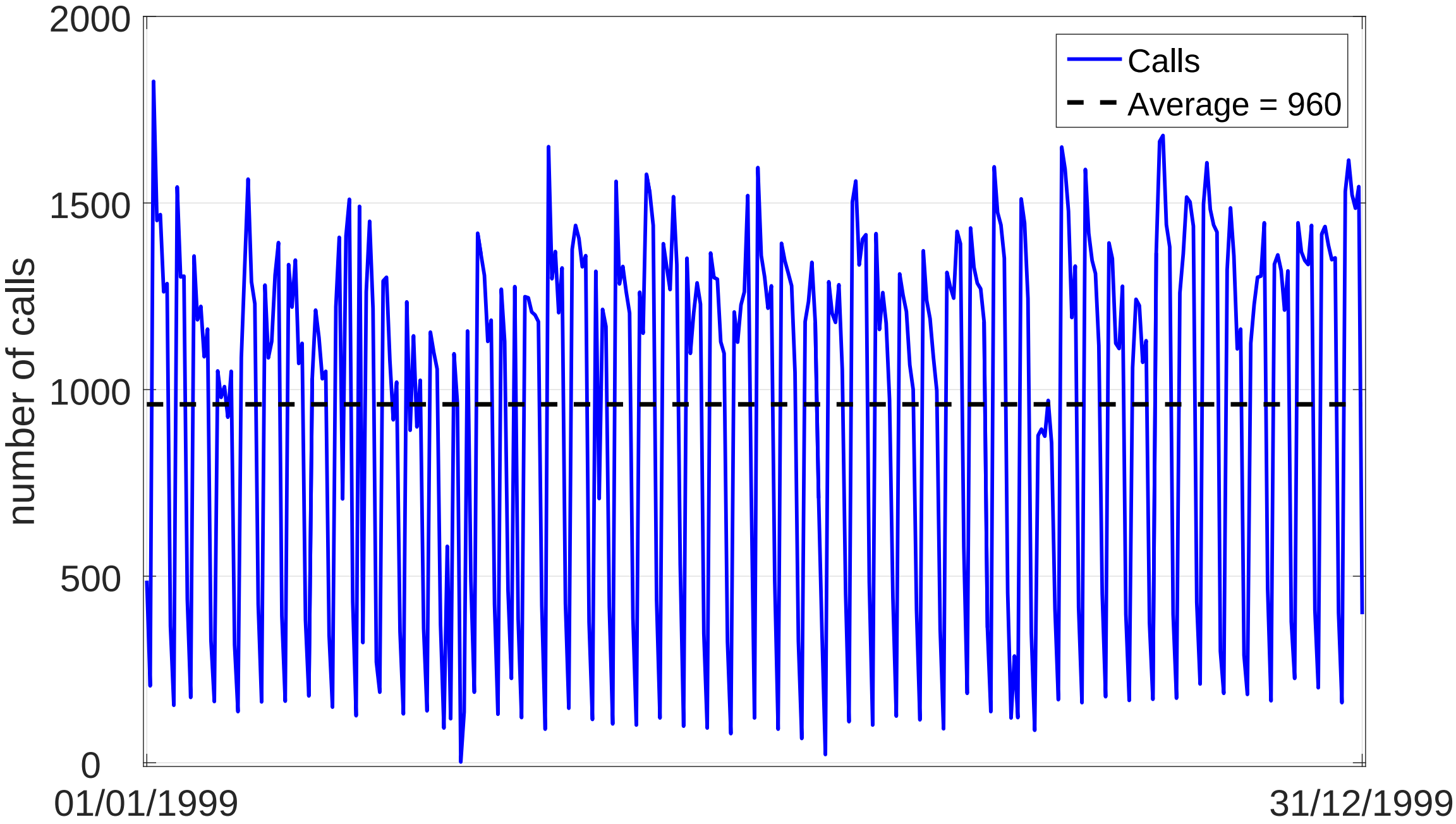

(a) Daily call arrivals

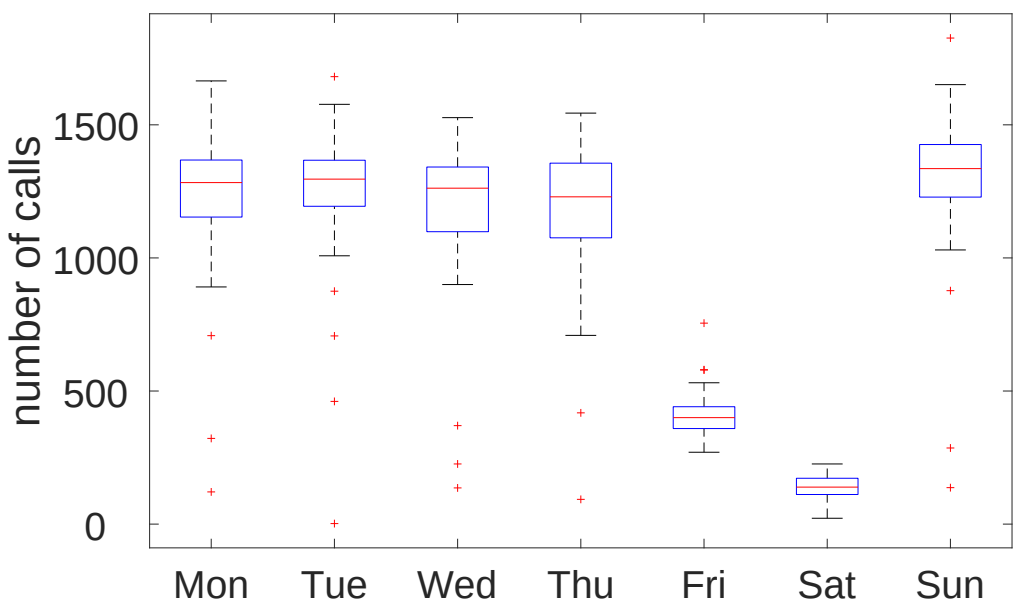

(b) Box plots of daily call arrivals per day for each day of the week.

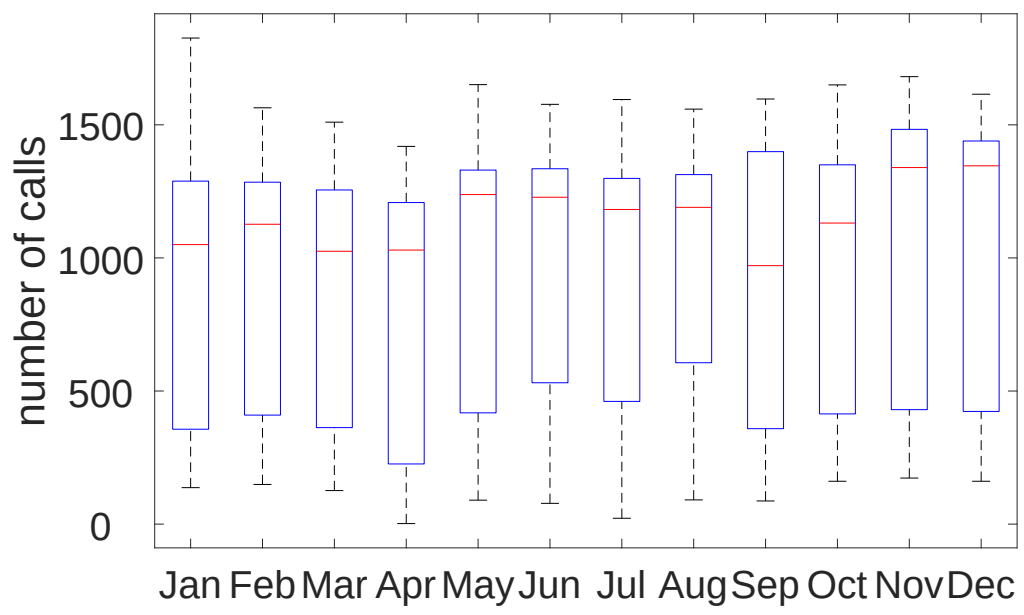

(c) Box plots of daily call arrivals per month for each month of the year.

*Notes:* panel (a) shows the daily call arrival series (continuous line) and its sample average (horizontal dashed line). In each box plot in panels (b) and (c), the central mark indicates the median, the bottom and top edges of the box indicate the 25th and 75th percentiles, respectively. The two bars, known as whiskers, are located at a distance of 1.5 times the interquartile range (i.e. the height of the box) below the 25th percentile and above the 75th percentile. Daily observations that fall above or below these two bars are denoted with crosses.

Figure B5: Daily call arrivals for the U.S. call center

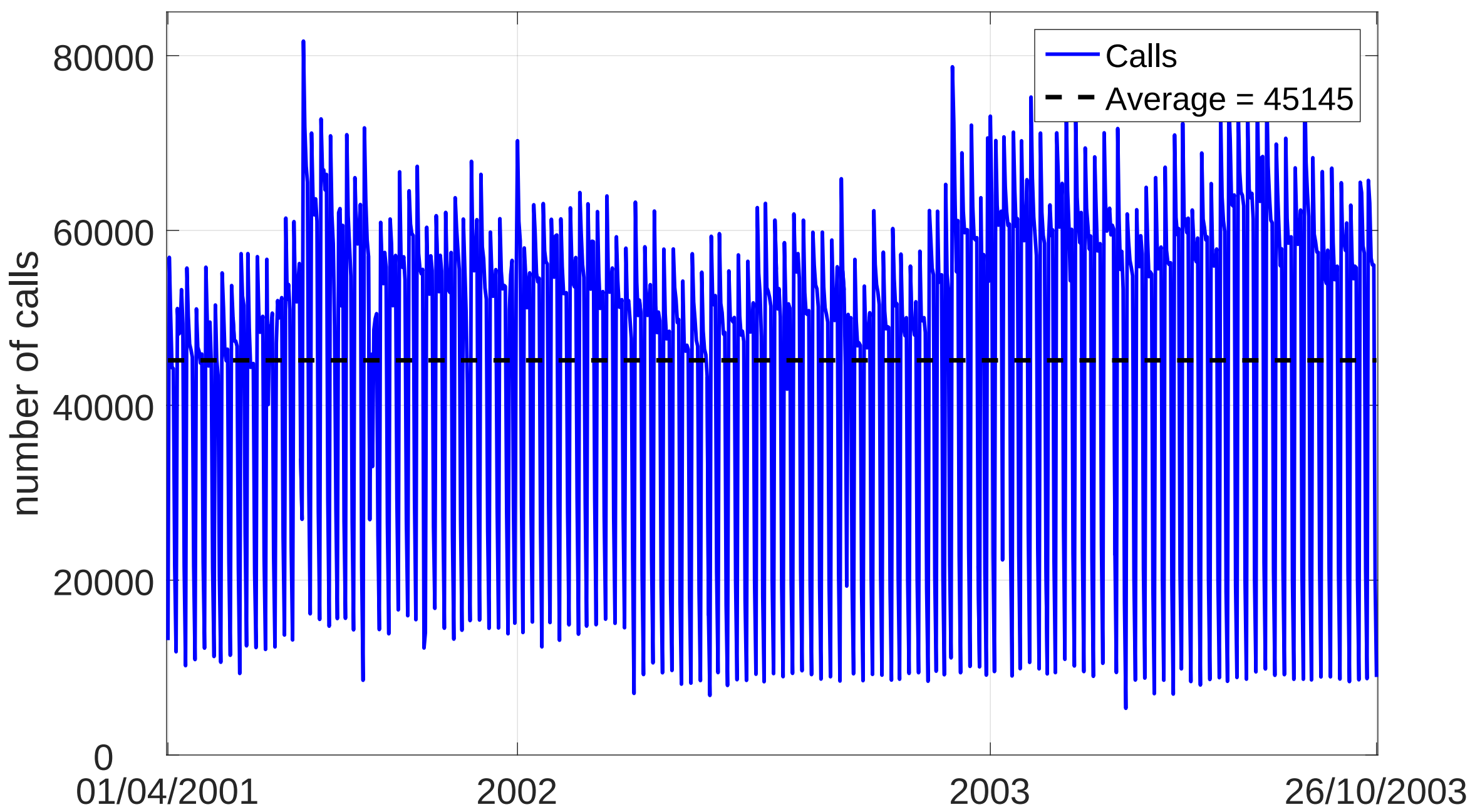

(a) Daily call arrivals

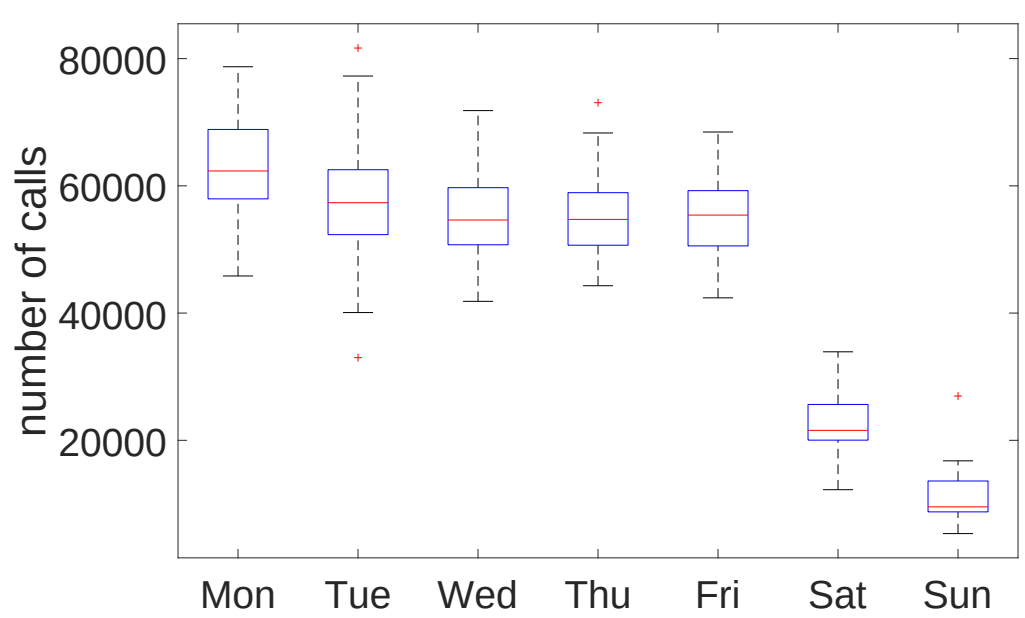

(b) Box plots of daily call arrivals per day for each day of the week.

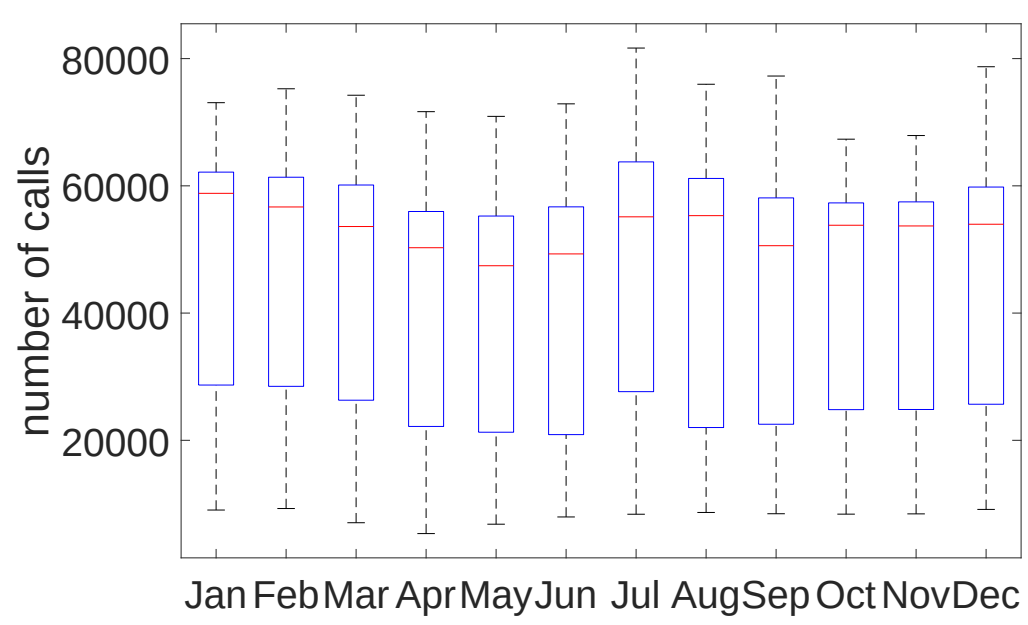

(c) Box plots of daily call arrivals per month for each month of the year.

*Notes:* panel (a) shows the daily call arrival series (continuous line) and its sample average (horizontal dashed line). In each box plot in panels (b) and (c), the central mark indicates the median, the bottom and top edges of the box indicate the 25th and 75th percentiles, respectively. The two bars, known as whiskers, are located at a distance of 1.5 times the interquartile range (i.e. the height of the box) below the 25th percentile and above the 75th percentile. Daily observations that fall above or below these two bars are denoted with crosses.

Figure B6: Log daily call arrivals and cubic spline fit

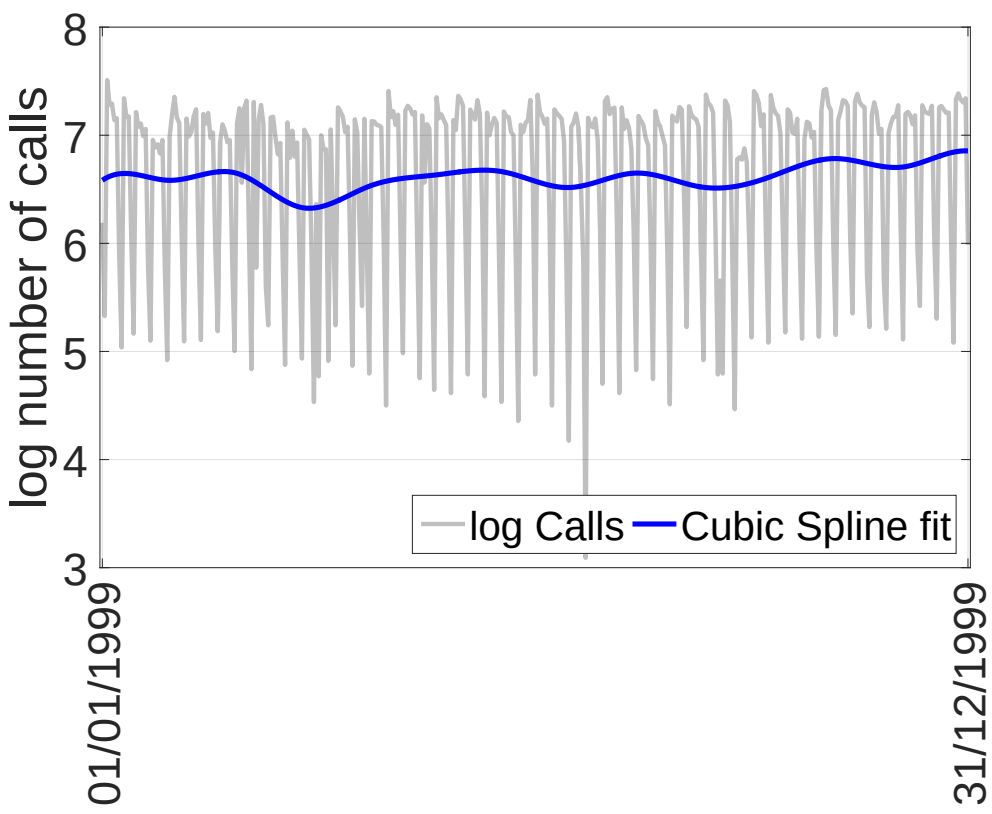

(a) Israel

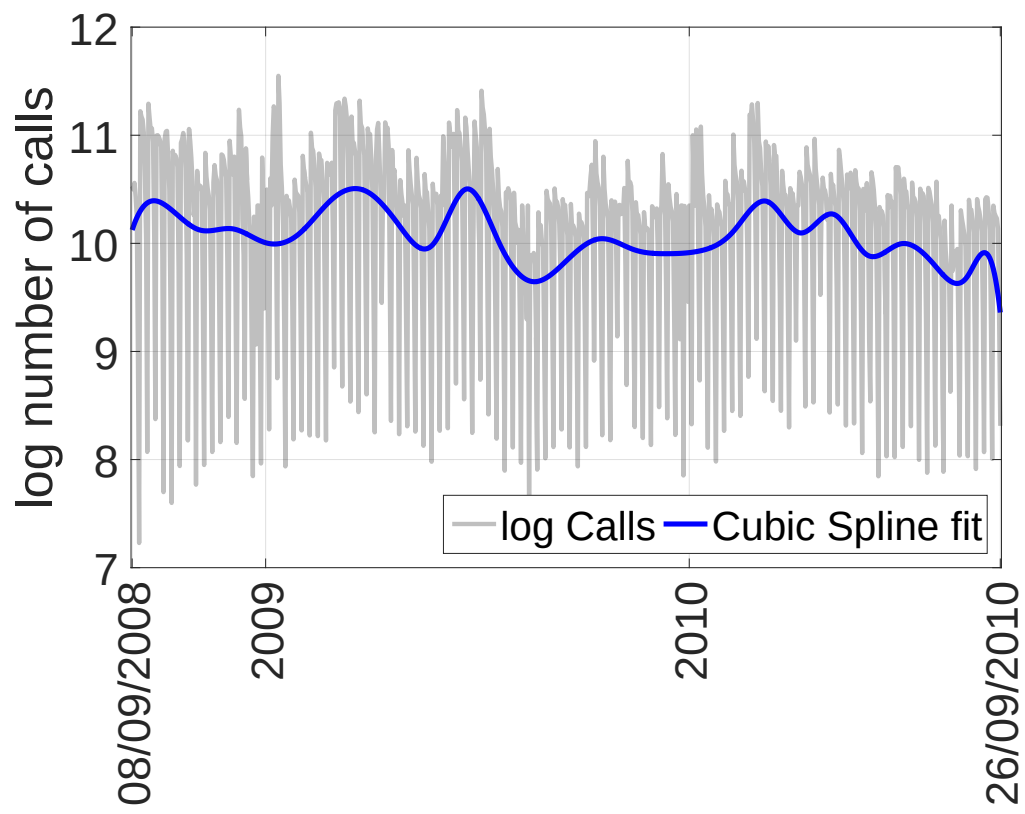

(b) Italy

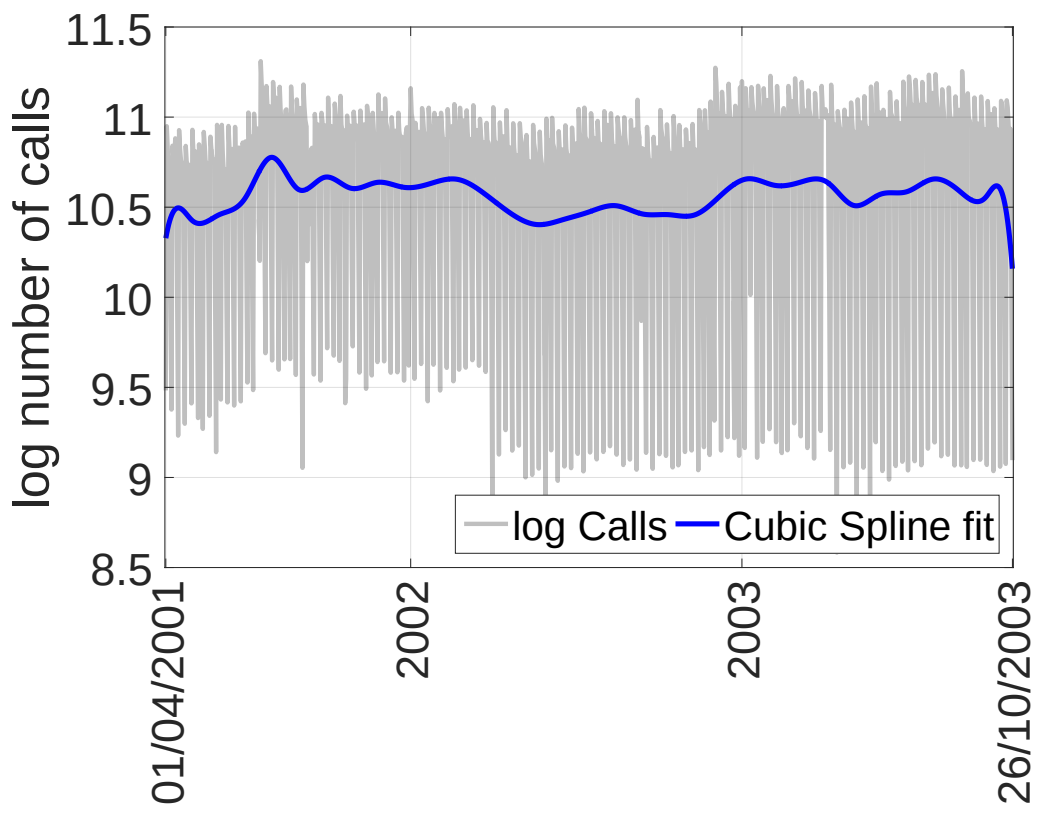

(c) US

*Notes:* for each call arrival series the figure shows the (log) number of incoming calls and the fitted value using a cubic spline. Closing days, if present, are set equal to the sample average before log-transforming the series.

Figure B7: Estimated autocorrelation of daily call arrivals

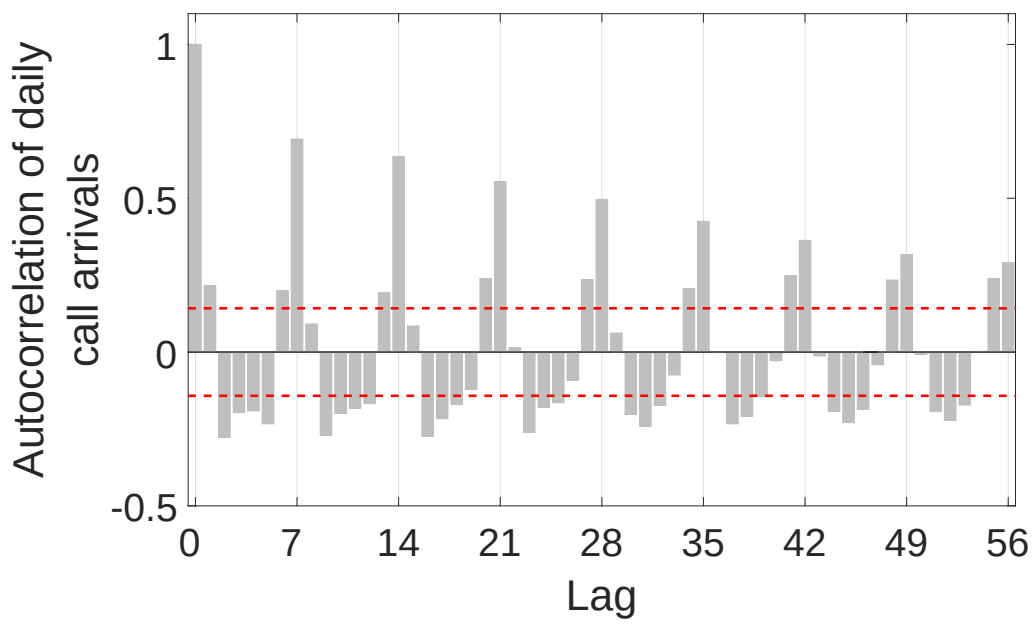

(a) Israeli call center.

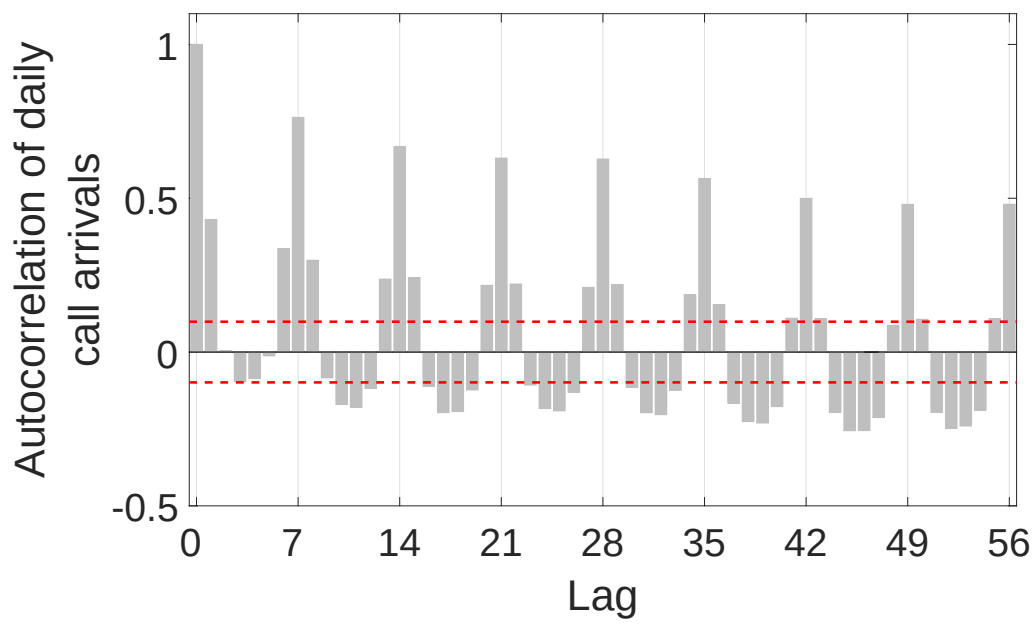

(b) Italian call center.

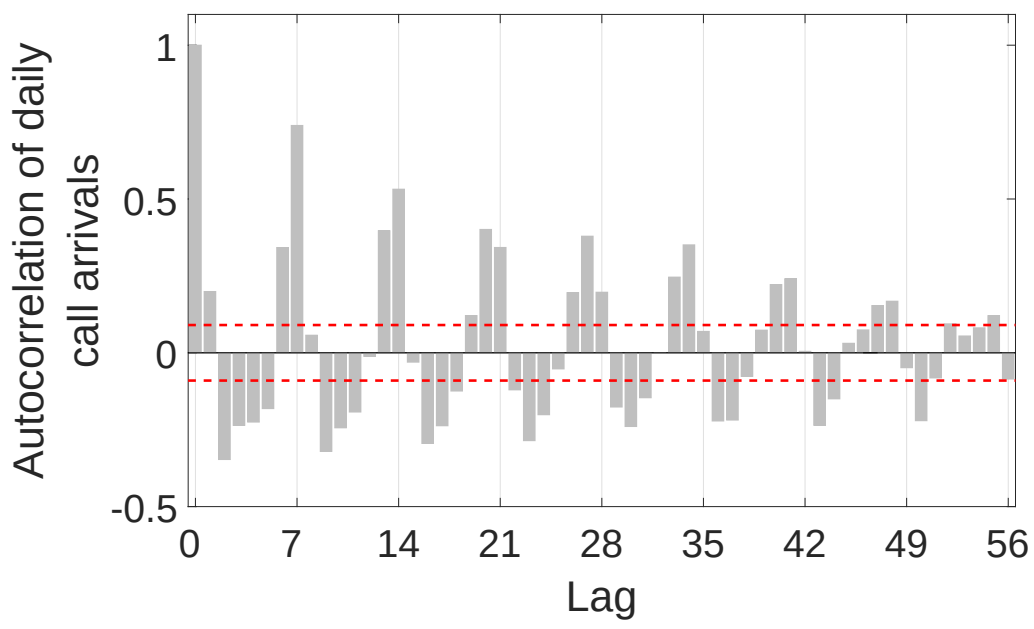

(c) US call center.

*Notes:* dashed lines are 95% confidence bands based on standard error calculated as the inverse of the square root of the number of observations.

## B.2 ARCH effects and seasonality in variance

The presence of conditional heteroskedasticity and seasonality in the variance of daily call arrivals is investigated in Table B2. For the ARMAX and SARMAX models we present a Lagrange Multiplier test of the null hypothesis of conditional heteroskedasticity, carried out as a test for the presence of serial correlation of order one in the squared residuals. Similarly, the test of the null hypothesis of seasonality in the variance is implemented by estimating an auxiliary regression of the squared residuals (from the ARMAX model or the SARMAX model) on a constant and six day-of-week dummy variables. The $p$-values are based on the $F$-test of the joint significance of the six dummy variables (see Franses and Paap, 2004).

Table B2: ARCH effects and seasonality in variance

| | | Israel | Italy | US |
|---|---|---|---|---|
| Model | Test | $p$-value | $p$-value | $p$-value |
| ARMAX | ARCH effects | 0.9939 | 0.0000 | 0.0000 |
| | Seasonality in variance | 0.0928 | 0.2270 | 0.0045 |
| SARMAX | ARCH effects | 0.9651 | 0.0000 | 0.0000 |
| | Seasonality in variance | 0.0882 | 0.2228 | 0.0049 |

*Notes:* For the ARMAX and the SARMAX models we present a Lagrange Multiplier test of the null hypothesis of conditional heteroskedasticity, carried out as a test for the presence of serial correlation of order one in the squared residuals. Similarly, the test of the null hypothesis of seasonality in the variance is implemented by estimating an auxiliary regression of the squared residuals (from the ARMAX model or the SARMAX model) on a constant and six day-of-week dummy variables. The $p$-values are based on an $F$-test of the joint significance of the six dummy variables (see Franses and Paap, 2004). The tests are carried out on the first estimation sample, that is on the first $\lfloor 0.55 \times T \rfloor$ observations, where $\lfloor \cdot \rfloor$ denotes the floor function.

## B.3 Overdispersion test

Call arrival volume represents a count and are often modeled as a Poisson process with time-varying arrival rate, $\mu_t$. The underlying assumption is that there is large population of potential customers, each of whom acts independently making calls with a very low probability (Ibrahim et al., 2016). Then, the total number of incoming calls in a given time period is approximately Poisson: $Y_t \sim Pois(\mu_t)$. In Poisson regression models the standard approach is to rely on $\mu_t = \exp(\mathbf{x}_t'\boldsymbol{\beta})$, where $\mathbf{x}_t$ is a vector of regressors (i.e. in this case, day-of-the-week dummies and the lagged call volume) and $\boldsymbol{\beta}$ is a vector of parameters. The equidispersion property is implicit in the Poisson approach: the variance of the call arrival volume in each period is equal to its expected value during the same time frame. This property is often not consistent with the features of call arrival data, that could well be over-dispersed (i.e. the variance is larger than the mean). Since Jung and Tremayne (2011) have shown that when forecasting with count data a dominating modeling approach does not exist, we consider three specifications based on the Exponential, Poisson and Negative Binomial distributions, respectively.

While the Poisson modeling approach posits that the arrival rate, $\mu_t$, is a deterministic function of the regressors, we can deal with overdispersion by assuming the arrival rate to be stochastic. In particular, we consider a Poisson distribution with a Gamma-distributed arrival rate, which implies that the series of call arrivals follows a Negative Binomial distribution. Assuming $\lambda_t = \mu_t \nu_t = \exp(\mathbf{x}_t' \boldsymbol{\beta})\nu_t$, where $\nu_t > 0$ is Gamma-distributed with unit mean and variance $\alpha$, $Y_t$ then follows a Negative Binomial distribution with mean $\mu_t$ and variance equal to $\mu_t(1 + \alpha\mu_t)$. Notice that the Negative Binomial distribution includes the Poisson as a special case when $\alpha = 0$. In fact, $\alpha$ controls the degree of overdispersion: the variance exceeds the mean when $\mu_t > 0$ and $\alpha > 0$, while for $\alpha = 0$, $E(Y_t|\mathbf{x}_t) = var(Y_t|\mathbf{x}_t) = \mu_t$.

Following Cameron and Trivedi (1990) a statistical test of overdispersion can be constructed after estimating the Poisson model. The starting point is to specify overdispersion of the form $var(Y_t|\mathbf{x}_t) = \mu_t(1+\alpha\mu_t)$. Notice that the previous expression corresponds to the variance of $Y_t$ when it follows a Negative Binomial distribution. A test statistic for $H_0 : \alpha = 0$ against $H_1 : \alpha \neq 0$ can be computed by estimating the Poisson model, constructing the fitted values $\hat{\mu}_t = \exp(\mathbf{x}_t'\hat{\boldsymbol{\beta}})$, the residuals $\hat{e}_t = Y_t - \hat{\mu}_t$ and running the following auxiliary OLS regression:

$$\hat{e}_t^2 - Y_t = \alpha\hat{\mu}_t^2 + \varepsilon_t \tag{10}$$

where $\varepsilon_t$ is an error term. We have implemented this test to check if the three call arrival series we have considered in the paper are overdispersed. The second column of Table B3 presents the estimates of the overdispersion parameter from running the auxiliary regression in equation (10). A low $p$-value in the third column denotes rejection of $H_0 : \alpha = 0$ and hence provides evidence against equidispersion of the data. The fourth column presents estimates of the degree of overdispersion implied by the Negative Binomial model. A low $p$-value in the fifth column denotes rejection of $H_0 : \alpha = 0$ and hence provides evidence against equidispersion of the data. The Poisson model underlying the auxiliary regression of the overdispersion test and the Negative Binomial model include a day-of-the-week dummies and the lagged dependent variable.

Table B3: Overdispersion test and estimates of the Negative Binomial model

| | Overdispersion test | | Negative Binomial model | |
|---|---|---|---|---|
| country | $\hat{\alpha}$ | $p$-value | $\hat{\alpha}$ | $p$-value |
| Israel | 0.0338 | (0.0000) | 0.0967 | (0.0000) |
| Italy | 0.0550 | (0.0000) | 0.5509 | (0.0000) |
| US | 0.0108 | (0.0000) | 0.0204 | (0.0000) |

*Notes:* the second column presents the estimates of the overdispersion parameter from running the auxiliary regression in equation (10). A low $p$-value in the third column denotes rejection of $H_0 : \alpha = 0$ and hence provides evidence against equidispersion of the data. The fourth column presents estimates of the degree of overdispersion implied by the Negative Binomial model. A low $p$-value in the fifth column denotes rejection of $H_0 : \alpha = 0$ and hence provides evidence against equidispersion of the data. The Poisson model underlying the auxiliary regression of the overdispersion test and the Negative Binomial model include a day-of-the-week dummies and the lagged dependent variable.